\documentclass{pasj00}
\draft

\begin{document}
\SetRunningHead{Y. Takeda et al.}{Non-LTE Line Formation of S~{\sc i} and Zn~{\sc i} in F--K Stars}
\Received{2005/03/23}
\Accepted{2005/07/12}

\title{Non-LTE Line-Formation and Abundances of Sulfur and Zinc 
in F, G, and K Stars 
\thanks{The electronic tables (E1, E2, E3, and E4) will be made available 
at the E-PASJ web site upon publication, while they are provisionally 
accessible at the WWW site of
$\langle$http://optik2.mtk.nao.ac.jp/$\ \widetilde{ }\ $takeda/sznabund/$\rangle$.}
}

%

\author{
Yoichi \textsc{Takeda,}\altaffilmark{1}
Osamu \textsc{Hashimoto,}\altaffilmark{2}
Hikaru \textsc{Taguchi,}\altaffilmark{2}
Kazuo \textsc{Yoshioka,}\altaffilmark{3}\\
Masahide \textsc{Takada-Hidai,}\altaffilmark{4}
Yuji \textsc{Saito,}\altaffilmark{5}
and 
Satoshi \textsc{Honda}\altaffilmark{1}}

\altaffiltext{1}{National Astronomical Observatory, 2-21-1 Osawa, Mitaka, Tokyo 181-8588}
\email{takedayi@cc.nao.ac.jp}
\altaffiltext{2}{Gunma Astronomical Observatory, 6860-86 Nakayama,
Takayama-mura, Agatsuma-gun, Gunma 377-0702}
\altaffiltext{3}{Gunma Study Center, The University of the Air, 
1-13-2 Wakamiya-cho, Maebashi, Gunma 371-0032}
\altaffiltext{4}{Liberal Arts Education Center, Tokai University, 
1117 Kitakaname, Hiratsuka, Kanagawa 259-1292}
\altaffiltext{5}{Department of Physics, Faculty of Science, 
Graduate School of Tokai University, 1117 Kitakaname, \\
Hiratsuka, Kanagawa 259-1292}


\maketitle

\begin{abstract}
Extensive statistical-equilibrium calculations on neutral
sulfur and zinc were carried out, in order to investigate how 
the non-LTE effect plays a role in the determination of S and Zn 
abundances in F, G, and K stars. 
Having checked on the spectra of representative F-type stars 
(Polaris, Procyon, and $\alpha$ Per) and the Sun that our 
non-LTE corrections yield a reasonable consistency between 
the abundances derived from different lines, 
we tried an extensive non-LTE reanalysis of published equivalent-width 
data of S~{\sc i} and Zn~{\sc i} lines for metal-poor halo/disk stars.  
According to our calculations, S~{\sc i} 9212/9228/9237 lines suffer 
significant negative non-LTE corrections amounting to $\ltsim$ 0.2--0.3 dex, 
while LTE is practically valid for S~{\sc i} 8683/8694 lines. 
Embarrassingly, as far as the very metal-poor regime is concerned, 
a marked discordance is observed between the [S/Fe] values from 
these two abundance indicators, in the sense that the former attains 
a nearly flat plateau (or even a slight downward bending) while the latter 
shows an ever-increasing trend with a further lowering of metallicity.
The reason for this discrepancy is  yet to be clarified.
Regarding Zn, we almost confirmed the characteristic tendencies 
of [Zn/Fe] reported from recent LTE studies (i.e., an evident/slight 
increase of [Zn/Fe] with a decrease of [Fe/H] for very metal-poor/disk 
stars), since the non-LTE corrections for the Zn~{\sc i} 4722/4810 
and 6362 lines (tending to be positive and gradually increasing 
towards lower [Fe/H]) are quantitatively of less significance 
($\ltsim$ 0.1 dex).
\end{abstract}

%


\section{Introduction}

The subject of this paper is to investigate how the non-LTE effect
influences the spectroscopic determination of sulfur and zinc 
abundances in F-, G-, and K-type stars (especially metal-poor ones) 
used for studying the Galactic chemical evolution history.

\subsection{Astrophysical Significance of S and Zn}

Sulfur belongs to the group of ``$\alpha$-capture elements''
(along with O, Mg, Si, Ca, and Ti), a large fraction of which are 
considered to have been synthesized in short-lived massive stars 
(and thrown out by type II supernovae) at the early-time of the Galaxy.
On the other hand, while the production of zinc is still controversial
and several possibilities are discussed (see, e.g., a summary 
presented by Chen et al. 2004), it is often regarded as being associated 
with ``Fe-group elements'' which were significantly produced in the 
later stage of the Galactic history by type Ia supernovae of 
longer-lived intermediate-mass stars.

The particularly important characteristic of these two elements 
is that they are chemically ``volatile'' such as the cases of 
C, N, and O. That is, owing to their low condensation temperature 
($T_{\rm c} \sim$ 650 K), they are difficult to condense into solids, 
unlike other ``refractory'' elements (Mg, Si, Fe-group elements, etc.) 
with a high $T_{\rm c}$ of $\sim$ 1300--1500 K. Namely, S and Zn are 
considered to be hardly affected by depletion due to dust formation. 
This fact is especially significant in the chemical composition of 
intergalactic gas, where these two volatile elements are likely to 
retain their original composition even for such a condition, while other 
refractory species (such as Mg or Fe) may have been significantly 
fractionated onto dust and depleted. For this reason, in the analysis 
of damped Lyman $\alpha$ (DLA) system of QSO absorption lines, 
S and Zn are generally regarded as being (depletion-independent) 
important tracers of the $\alpha$ group and the Fe group, respectively, 
which provides us a possibility to use [S/Zn] and [Zn/H] determined 
from DLA as a ``chemical clock'' of high-$z$ universe (see, e.g., 
the summary of Nissen et al. 2004 and the references therein).
Anyway, as a first step toward such an advanced application, 
the nucleosynthesis history of these elements in our Galaxy has to 
be well understood by observationally establishing the behavior of 
[S/Fe] and [Zn/Fe] with a change of [Fe/H] in metal-poor stars. 
Yet, this problem has not necessarily been straightforward. 
Especially, we are still in a confusing situation concerning the case of S.

\subsection{Controversy over the Behavior of [S/Fe]}

Since sufficiently strong sulfur lines (measurable even in late-type 
metal-poor stars) are located in the near-IR region, earnest investigations
on the [S/Fe] vs. [Fe/H] relation began in the 1980's when efficient 
solid-state detectors had become widely used. Further, these early studies
based on S~{\sc i} 8693/8694 lines sufficed to reveal the gradual 
increase of [S/Fe] from $\sim 0$ (at [Fe/H] $\sim 0$) to $\sim +0.5$ 
(at [Fe/H] $\sim -1.5$) with a decrease of metallicity (Clegg et al. 
1981; Fran\c{c}ois 1987, 1988), which is typical for the $\alpha$ group.

The more important issue is, however, how it behaves itself in the 
very metal-poor regime of $-3 \ltsim$ [Fe/H] $\ltsim -2$.
The first two trials of investigating [S/Fe] in this region based on
the S~{\sc i} 8683/8684 lines (being very weak) resulted in 
a conclusion that [S/Fe] continues to ever increase with a further 
lowering of [Fe/H] (up to [S/Fe] $\sim +1$ at [Fe/H] $\sim -3$), 
while retaining its slope almost unchanged (Israelian, Rebolo 2001;
Takada-Hidai et al. 2002). This makes a marked contrast with
other $\alpha$-group elements (e.g., Mg, Si, Ca, Ti) showing a ``flat'' 
tendency (i.e., [$\alpha$/Fe] attains an almost constant value
below [Fe/H] $\ltsim -1$).
Thus, these authors suggested that [S/Fe] (along with [O/Fe]
\footnote{Oxygen is also suspected to show such an ever-increasing 
trend  though it still remains as a controversial matter
(i.e., different behaviors are suggested from different lines; 
cf. Takeda 2003 and the references therein).}) does not conform to 
the standard $\alpha$ behavior, and that an adequate chemical
evolution model (e.g., hypernovae or time-delayed deposition) 
reasonably reproducing such a tendency may be required.

Then, the situation turned into a new aspect, when the 
S~{\sc i} lines of the 9212/9228/9237 triplet (much stronger 
than S~{\sc i} 8693/8694 lines) began to be used for exploring 
the sulfur abundances of very metal-deficient stars based on the two
recent studies of Ryde and Lambert (2004) and Nissen et al. (2004).
Interestingly, they both arrived at the same conclusion 
that [S/Fe] attains an approximately constant value of $\sim +0.3$ 
at [Fe/H] $\ltsim -1$, which is nothing but a typical $\alpha$-group
trend explainable with a standard chemical evolution model.
Namely, their results derived from the 9212/9228/9237 lines are 
evidently opposed to what Israelian and Rebolo (2001) and 
Takada-Hidai et al. (2002) reported based on the 8693/8694 lines. 
The reason for this discrepancy should be clarified by all means.

Here, one point should be mentioned. Their analyses are based on 
the belief that reliable abundances can be obtained from the 
9212/9228/9237 lines on the assumption of LTE.
It appears, however, that their arguments are a kind of 
speculation, lacking any firm background; namely, they seem 
to overinterpret the consequence of Takada-Hidai et al. (2002)
(i.e., non-LTE corrections are negligible for the 8693/8694 lines 
of multiplet 6) and extend it to the 9212/9228/9237 lines of
multiplet 1, without paying attention to the large difference 
in the line-formation mechanism between these two multiplets.  
Therefore, in order to examine the validity of their treatment, an 
extensive study has to be carried out on how and to what extent 
the non-LTE effect is important for these multiplet 1 sulfur lines 
over a wide range of stellar parameters.

In addition, a comprehensive comparison study for the lines of both 
multiplets (1 and 6) on a data sample comprising as many stars as possible 
is necessary. Takada-Hidai et al. (2005) very recently published 
their new observations of both 8693/8694 and 9212/9228/9237 lines 
for some $\sim 20$ stars carried out at Okayama Astrophysical 
Observatory, and made an effort to investigate how the abundances 
derived from these two different multiplets compare with each other, 
while invoking the assumption of LTE as was done by Ryde and Lambert 
(2004) and Nissen et al. (2004).
Unfortunately, however, the insufficient number of data at 
[Fe/H] $\ltsim -2$ in their sample, coupled with a large scatter, did 
not allow any convincing argument on the [S/Fe] vs. [Fe/H] relationship 
of metal-poor stars or on the validity of LTE (cf. their figure 3). 
This lesson actually taught us the necessity of carrying out an 
extensive non-LTE reanalysis on almost all available observational 
data published so far combined together, which is thus the first 
purpose of this investigation.

\subsection{Does Zn Scale with Fe?}

The run of [Zn/Fe] with [Fe/H] in Galactic metal-poor stars is 
not as controversial as the case of sulfur. Nevertheless, 
some problematic points have emerged from recent studies.

Since the pioneering work of Sneden and Crocker (1988) and
Sneden, Gratton, and Crocker (1991), it has long been believed that Zn 
practically scales with Fe (i.e.,  [Zn/Fe] $\simeq 0$) over a wide range of 
metallicities ($-3 \ltsim$ [Fe/H] $\ltsim 0$). As a matter of fact, 
this is the reason why Zn is regarded as a substitute for Fe in the 
condition where the refractory Fe may be appreciably depleted 
(e.g., analysis of DLA line system mentioned in subsection 1.1).

However, recent studies extending to an ultra-metal-poor region, or those 
with higher accuracy than before, have gradually cast doubt on this 
simple picture. For example, at metallicities lower than [Fe/H] $\ltsim -3$,
it has been reported that [Zn/Fe]  progressively increases with
a decrease in [Fe/H] (cf. Primas et al. 2000; Cayrel et al. 2004).
Also, according to Chen, Nissen, and Zhao (2004) investigation on the Zn 
abundances of Galactic disk stars based on the weaker Zn~{\sc i} 6362 line,
which may be more reliable than the usually used 4722/4810 lines,
[Zn/Fe] appears to show a slightly increasing tendency as [Fe/H]
decreases from $\sim 0$ to $\sim -0.6$, consistently with the
findings of Reddy et al. (2003) and Bensby, Feltzing, and Lundstr\"{o}m (2003).
Bensby, Feltzing, and Lundstr\"{o}m (2003) also suggested a tendency of 
``uprising'' [Zn/Fe] at [Fe/H] $>0$ as well as a separation of [Zn/Fe] 
between thick-disk and thin-disk stars.
Furthermore, the [Zn/Fe] vs. [Fe/H] relation constructed from published 
results (combined) appear to suggest a shallow/marginal dip at 
[Fe/H] $\sim -1.2$ (cf. figure 8 of Mishenina e al. 2002 or figure 7 
of Nissen et al. 2004; though the number of stars is rather insufficient 
around this metallicity.) 
In any case, there are now several good reasons to suspect that the real 
situation is not so simple (i.e., scaling with Fe) as has ever 
been considered. Hence, it is needed to establish the trend of [Zn/Fe] 
as a function of metallicity as precisely as possible. 

Here, it is one of our concerns that all of these studies were done 
under the assumption of LTE. As a matter of fact, no investigation
has so far been carried out on the non-LTE effect of zinc, so far as
we know. Then, do some of the departures from [Zn/Fe] $\simeq 0$
described above have something to do with neglecting the non-LTE 
effect? Or alternatively, is the non-LTE correction so negligible 
as to guarantee the reliability of such trends derived from LTE? 
In search of the answers to these questions, we decided to carry out
statistical equilibrium calculations on neutral zinc, in order to 
estimate the non-LTE abundance corrections for Zn~{\sc i} 
4722/4810/6362 lines for a wide range of stellar parameters, and 
to perform an extensive non-LTE reanalysis of published equivalent-width 
data of these Zn~{\sc i} lines toward establishing the [Zn/Fe] vs. 
[Fe/H] relation in our manner. This is the second aim of this study.

\section{Non-LTE Line Formation and Abundance Corrections}

\subsection{Sulfur}

\subsubsection{Non-LTE calculations on S I atom}

The procedures of our statistical-equilibrium calculations for 
neutral sulfur are almost the same as those described in Takada-Hidai 
and Takeda (1996) and Takada-Hidai et al. (2002). For the present study, 
however, we reconstructed a new S~{\sc i} model atom comprising 57 terms 
(up to 3s$^{2}$3p$^{3}$8f~$^{3}$F at 81837 cm$^{-1}$)
and 191 transitions, while using Kurucz and Bell's (1995) compilation 
of atomic data ($gf$ values, levels, etc.), which we believe to be 
more realistic than the previous one (56 terms and 173 transitions)
based basically on Kurucz and Peytremann's (1975) data.

The treatment of the photoionization cross sections is the same as 
described in subsection 3.3 of Takada-Hidai and Takeda (1996);
namely, the available cross-section values compiled by Mathisen (1984) 
were adopted for the lowest three terms (original sources: 
Tondello 1972 for 3p$^{4}$~$^{3}$P; Chapman, Henry 1971 for 
3p$^{4}$~$^{1}$D; McGuire 1979 for 3p$^{4}$~$^{1}$S), while 
the hydrogenic approximation was assumed for the remaining terms. 

Regarding the collision cross section, we followed the recipe
adopted in subsubsection 3.1.3 of Takeda (1991).
It should also be mentioned that no correction [i.e., a correction 
factor of $k=1$ or logarithmic correction of $h (\equiv \log k) = 0$] 
was applied to the H~{\sc i} collision rates computed with the classical 
approximate formula (Steenbock, Holweger 1984; Takeda 1991), 
though test calculations with varying $k$ from 10 to $10^{-3}$ were 
also performed (see below). The validity of this choice is discussed
in connection with the analysis of bright F stars presented in section 3.

Since we planned to make our calculations applicable to stars 
from metal-rich  (population I) down to very low 
metallicity (extreme population II) at early-F through early-K 
spectral types in various evolutionary stages 
(i.e., dwarfs, subgiants, giants, and supergiants),
we carried out non-LTE calculations on an extensive grid of 
210 ($6 \times 5 \times 7$) model atmospheres, resulting from 
combinations of six $T_{\rm eff}$ values 
(4500, 5000, 5500, 6000, 6500, 7000 K), five $\log g$ values 
(1.0, 2.0, 3.0, 4.0, 5.0), and seven metallicities (represented by [Fe/H])
(+0.5, 0.0, $-0.5$, $-1.0$, $-2.0$, $-3.0$, $-4.0$).
As for the stellar model atmospheres, we adopted Kurucz's (1993) ATLAS9 
models corresponding to a microturbulent velocity ($\xi$) of 2~km~s$^{-1}$.
Regarding the sulfur abundance used as an input value in non-LTE 
calculations, we assumed 
$A_{\rm S}^{\rm input}$ = 7.21 + [Fe/H] + [S/Fe], 
where the values of 0.0 (for [Fe/H] = +0.5, 0.0, $-0.5$)
and +0.5 (for [Fe/H] = $-1.0$, $-2.0$, $-3.0$, $-4.0$) 
were assigned to [S/Fe] in order to roughly simulate the
behavior of this ratio (cf. subsection 4.1).
The solar sulfur abundance of 7.21 was adopted from
Anders and Grevesse (1989) (which is used also in the ATLAS9 models). 
The microturbulent velocity (appearing in the 
line-opacity calculations along with the abundance) was assumed 
to be 2~km~s$^{-1}$, to make it consistent with the model atmosphere.
\subsubsection{Non-LTE characteristics of S I line formation}

In figure 1 are shown the $S_{\rm L}(\tau)/B(\tau)$ (the ratio of 
the line source function to the Planck function, and nearly equal to 
$\simeq b_{\rm u}/b_{\rm l}$, where $b_{\rm l}$ and $b_{\rm u}$ are 
the non-LTE departure coefficients for the lower and upper levels, 
respectively) and  $l_{0}^{\rm NLTE}(\tau)/l_{0}^{\rm LTE}(\tau)$ 
(the NLTE-to-LTE line-center opacity ratio, and nearly equal to 
$\simeq b_{\rm l}$) for each of the multiplet 1 (9212/9228/9237) and
multiplet 6 (8693/8694) transitions for a representative set of 
model atmospheres.
Two especially important characteristics can be read from this figure:\\
--- Generally, the inequality relations of $S_{\rm L}/B<1$ 
(dilution of line source function) and $l_{0}^{\rm NLTE}/l_{0}^{\rm LTE} >1$ 
(enhanced line-opacity) hold in the important line-forming region
for both cases of multiplets 1 and 6, which means that the non-LTE 
effect almost always acts in the direction of strengthening the 
9212/9228/9237 and 8693/8694 lines; i.e., the non-LTE correction is
generally negative.\\
--- The departure from LTE in the line opacity
(the enhancement of $l_{0}^{\rm NLTE}/l_{0}^{\rm LTE}$ over 1)
becomes prominent for higher $T_{\rm eff}$ and/or very low-metallicity 
case, which indicates that the non-LTE effect may become significant
in very low metallicity regime of [Fe/H] $\sim -3$ down to $-4$
even if the line-strength is weak.
\subsubsection{Grid of non-LTE corrections for S I lines}

Based on the results of these calculations, we computed extensive
grids of theoretical equivalent-widths and the corresponding non-LTE
corrections for the eight selected important lines (S~{\sc i} 8693 and 
8694 lines of multiplet 6; 9212, 9228, and 9237 lines of multiplet 1;
10455, 10456, and 10459 lines of multiplet 3) for each of the model 
atmospheres as follows.

For an assigned sulfur abundance ($A^{\rm a}$) and
microturbulence ($\xi^{\rm a}$), we first 
calculated the non-LTE equivalent width ($W^{\rm NLTE}$) of the line 
by using the computed non-LTE departure coefficients ($b$) for each model 
atmosphere. Next, the LTE ($A^{\rm L}$) and NLTE ($A^{\rm N}$) abundances
were computed from this $W^{\rm NLTE}$
while regarding it as if being a given observed equivalent width.
We could then obtain the non-LTE abundance correction, $\Delta$, which is 
defined in terms of these two abundances as 
$\Delta \equiv A^{\rm N} - A^{\rm L}$.

Strictly speaking, the departure coefficients [$b(\tau)$] for 
a model atmosphere correspond to the sulfur abundance and 
the microturbulence of $A_{\rm S}^{\rm input}$ and 
2~km~s$^{-1}$ adopted in the non-LTE calculations (cf. subsubsection 2.1.1). 
Nevertheless, considering the fact that the departure coefficients
(i.e., {\it ratios} of NLTE to LTE number populations) are
(unlike the population itself) not very sensitive to small changes in 
atmospheric parameters, we also applied such computed $b$ values to 
evaluating $\Delta$ for slightly different $A^{\rm a}$ and $\xi^{\rm a}$ 
from those fiducial values assumed in the statistical equilibrium 
calculations.
Hence, we evaluated $\Delta$ for three $A^{\rm a}$ values
($A_{\rm S}^{\rm input}$ and $\pm 0.3$ dex perturbations)
as well as three $\xi$ values (2~km~s$^{-1}$ and $\pm 1$~km~s$^{-1}$ 
perturbations) for a model atmosphere using the same departure coefficients.

We used the WIDTH9 program (Kurucz 1993) for calculating the equivalent
width for a given abundance, or inversely evaluating the abundance 
for an assigned equivalent width. Actually, this program was considerably 
modified in many respects: e.g., the treatment of a blended feature due to
multiplet components, the incorporation of the non-LTE departure 
in the line source function as well as in the line opacity, etc. 
The adopted line data ($gf$ values, damping constants, etc.) 
are given in table 1. 
Since the S~{\sc i} 9228 lines are located on the wing of Paschen line
(H~{\sc i} P$_{9}$ 9229.0), we replaced the hydrogen-line opacities of 
the original Kurucz's code [based on classical Griem's (1960, 1967) approximation] 
by a more updated one based on the extended VCS theory calculated 
by Lemke (1997).

As a demonstrative example of non-LTE corrections, we give 
the $\xi$ = 2~km~s$^{-1}$ results for the S~{\sc i}~8694 and 9212 lines 
computed for representative parameters in table 2, 
where we also present the cases of $h =+1, -1, -2, -3$ in addition to 
the fiducial $h = 0$ [$h$ ($\equiv \log k$) is the logarithmic H~{\sc i} 
collision correction to be applied to the classical formula; cf.
subsubsection 2.1.1) for comparison. As can be seen from table 2, 
the non-LTE effect becomes more appreciable with a decrease of $h$,
as expected. Also, conspicuously large (negative) non-LTE corrections 
(accompanied by large line-strengths) seen in low-gravity and/or 
high-$T_{\rm eff}$ stars are worth noting.

Since the S~{\sc i} photoionization cross sections that we adopted 
may not be sufficiently up to date (cf. subsubsection 2.1.1), 
we investigated how changing the cross-section values by factors of
0.1 and 10 would affect the non-LTE corrections; the results are
also given in table 2 ( $\delta(\Delta)_{-}$ and $\delta(\Delta)_{+}$ ).
While the resulting changes are not necessarily straightforward,
we can see that the extent of the negative corrections tends to be 
reduced (i.e., less negative) by increasing the photoionization,
which may be interpreted as that lines tend to be weakened 
(i.e., bringing the non-LTE correction in the positive direction)
by the photo-overionization effect. From a quantitative view, however, 
the variations in $\Delta$ are not significant in most cases. 
Yet, the exceptional cases are high-$T_{\rm eff}$ as well as low-$\log g$ 
stars, for which changes are appreciably large to be several-tenth dex
(reflecting the importance of photoionization by UV radiation).

The complete results of the non-LTE corrections (for all combinations 
of $T_{\rm eff}$, $\log g$, and $\xi$ values for each of the 8 S~{\sc i} lines, 
though only for the case of $h = 0$) are given in electronic table E1.

\subsection{Zinc}

\subsubsection{Non-LTE calculations on Zn I atom}

Again invoking Kurucz and Bell's (1995) compilation of atomic data,
we constructed a Zn~{\sc i} model atom consisting of 44 terms (up to 
3d$^{10}$4s\ 14d~$^{3}$D at 75112 cm$^{-1}$) and 87 transitions. 
The hydrogenic approximation was assumed for the photoionization
rates from all terms. Regarding the collisional rates (due to electron
and neutral hydrogen), we followed the classical formulae described 
in subsubsection 3.1.3 of Takeda (1991).
Although we eventually adopted $h = 0$ (i.e., without applying any 
correction to the classical value) for the H~{\sc i} collision rates,
the effect of varying $h$ was also examined (cf. table 3).

Similarly to the case of S~{\sc i}, we carried out extensive non-LTE
calculations on a grid of 210 model atmospheres.
The input Zn abundance in non-LTE calculations was assumed to be
$A_{\rm Zn}^{\rm input}$ = 4.60 + [Fe/H] + [Zn/Fe], 
where we assigned the values of 0.0 (for [Fe/H] = +0.5, 0.0, $-0.5$, 
$-1.0$, $-2.0$) and +0.5 (for [Fe/H] = $-3.0$, $-4.0$) 
to [Zn/Fe] while considering the recently observed supersolar
ratio at extremely low metallicities  (cf. subsection 4.2).
The solar zinc abundance of 4.60 was adopted from
Anders and Grevesse (1989).
\subsubsection{Non-LTE characteristics of Zn I line formation}

Similarly to figure 1, we show in figure 2 the behaviors of 
$S_{\rm L}(\tau)/B(\tau)$ and  $l_{0}^{\rm NLTE}(\tau)/l_{0}^{\rm LTE}(\tau)$ 
for two important Zn~{\sc i} transitions of multiplet 2 (4722/4810) and 
multiplet 6 (6362) for representative model atmospheres.
The noteworthy characteristics recognized from this figure are as follows:\\
--- Unlike the case of S~{\sc i}, $S_{\rm L}$ tends to be superthermal
($S_{L} > B$) at important line-forming regions, which generally acts
in the direction of line-weakening. \\
--- The behavior of the NLTE-to-LTE line-opacity ratio differs from 
case to case; it tends to be greater than unity (line-strengthening)
at lower $T_{\rm eff}$, while it becomes appreciably less than unity 
(line-weakening) at higher $T_{\rm eff}$ (especially for very 
metal-poor cases). Also, the trend of multiplet 2 is significantly 
different from the case of multiplet 6.\\
--- Combining these characteristics mentioned above, we can expect
that the net non-LTE effect is rather complicated, since two mechanisms 
may occasionally act in the opposite direction and compensate with 
each other. Yet, we may roughly state that the non-LTE effect on the 
Zn~{\sc i} lines is in many cases a slight line-weakening (i.e., positive 
non-LTE correction), though it may sometimes act to strengthen the line
case by case.
\subsubsection{Grid of non-LTE corrections for Zn I lines}

As was described in subsubsection 2.1.3 for the case of sulfur, we 
computed extensive grids of the theoretical equivalent-widths and 
the corresponding non-LTE corrections for the three important zinc 
lines (Zn~{\sc i} 4722 and 4810 lines of multiplet 2; 6362 line of 
multiplet 6) for each of the model atmospheres.
The adopted line data are given in table 1.

As demonstrative examples of non-LTE corrections, we give 
the $\xi$ = 2~km~s$^{-1}$ results for the Zn~{\sc i}~4810 and 6362 lines 
computed for representative parameters in table 3, 
where we also present the cases of $h =+1, -1, -2, -3$ in addition to 
the adopted case of $h = 0$. 
Again, it is apparent that the non-LTE effect becomes more appreciable 
with a decrease of $h$. 
We see, however, the extent of non-LTE correction is generally
small and comparatively insignificant; also, its sign becomes
positive as well as negative case by case. This can be understood
from the characteristics of non-LTE line-formation described in
subsubsection 2.2.2.
Yet, appreciably large positive non-LTE corrections amounting up to 
$\sim 0.3$~dex seen in low-gravity/high-$T_{\rm eff}$/low-[Fe/H] cases 
are worth noting.

As was done for sulfur, we also investigated how changing the Zn~{\sc i} 
photoionization cross sections affects the non-LTE corrections; 
the results are given table 3. While we can observe a roughly 
similar tendency to the case of S, the variations are quantitatively
insignificant ($\ltsim 0.1$ dex).

The complete results of the non-LTE corrections (for all combinations 
of $T_{\rm eff}$, $\log g$, $\xi$ values for each of the three 
Zn~{\sc i} lines, though only for the fiducial case of $h = 0$) 
are given in electronic table E2.

\section{Analysis of S and Zn Lines for Selected F-type Stars and the Sun}

\subsection{Importance of Abundance Consistency Check}

According to the calculations described in the previous section, unlike the case 
of zinc where the non-LTE effects on Zn~{\sc i} lines are comparatively
minor and insignificant, the extents of the non-LTE corrections for 
S~{\sc i} 9212/9228/9237 lines are appreciably large and of practical
importance (cf. table 2 in subsubsection 2.1.3 and electronic table E1). 
Thus, it is requisite to examine the validity of our calculations
(e.g., whether our choice of $h=0$ is reasonable or not) 
by empirically checking the abundance consistency between the S~{\sc i} 
8693/8694 and 9212/9228/9237 lines in actual stars. 

Towards this aim, we decided to carry out abundance analyses with 
respect to these S~{\sc i} lines based on our non-LTE calculations, 
while using the spectra of representative bright F-type stars: 
$\alpha$ Per (F5~Ib), Polaris (F8~Ib), and Procyon (F5~IV) [along with 
the solar flux spectra of Kurucz et al. (1984)]. 
The reason why we selected these F stars with $T_{\rm eff} \gtsim 6000$~K 
is that such stars (especially low-gravity supergiants) may serve as 
the most suitable touchstone for this purpose, since the non-LTE effect 
on these S~{\sc i} lines becomes so large that their sensitivity to 
changing $h$ may be quantitatively appreciable in such a condition of 
higher $T_{\rm eff}$ and lower $\log g$ (i.e., the case where lines 
are strong). 

In addition, we also analyzed the Zn~{\sc i} 
4722/4810 and 6362 lines of these F stars and the Sun, though their 
non-LTE corrections are so small that we can not say much about 
the adequacy of our non-LTE calculations.

\subsection{Observational Data}

The observations of these three target stars ($\alpha$ Per, Polaris, 
and Procyon) were carried out by using the new high-dispersion 
echelle spectrograph GAOES (Gunma Astronomical Observatory Echelle 
Spectrograph), which was recently installed at the Nasmyth Focus of 
the 1.5~m reflector of the Gunma Astronomical Observatory
and can obtain spectra of high wavelength resolution ($R \sim 70000$ 
for the standard slit width of $1''$) along with a wide 
wavelength coverage ($\sim 1800$~$\rm\AA$ by using the 2K$\times$4K CCD).
[See Hashimoto et al. (2002, 2005) for more information.]

For each star, we obtained spectra at three wavelength regions
(region G: 4600--6400~$\rm\AA$; region R: 5900--7600~$\rm\AA$;
region I: 7600--9350~$\rm\AA$). Most of the data were obtained
in the observing period of 2004 December 14--17, except for the region G
(2004 August 31) and region I (2004 October 27) spectra of $\alpha$ Per.

The data reduction (bias subtraction, flat-fielding, 
aperture-determination, scattered-light subtraction, 
spectrum extraction, wavelength calibration, and continuum-normalization) 
was performed using the ``echelle'' package of IRAF.\footnote{IRAF is 
distributed by the National Optical Astronomy Observatories, 
which is operated by the Association of Universities for  Research  
in Astronomy, Inc., under cooperative agreement with the National 
Science Foundation.}

Since the spectrum portion including S~{\sc i} 9212/9228/9237 lines 
contains numerous H$_{2}$O lines originating from Earth's atmosphere, 
it was divided by the spectrum of $\gamma$ Cas (rapid rotator) by using 
the IRAF task ``telluric'' to remove these telluric lines, which turned
out to be reasonably successful in most cases.
The final spectra of three stars (along with the solar flux spectrum 
for comparison) at the wavelengths corresponding to the relevant
S~{\sc i} and Zn~{\sc i} lines are shown in figure 3.

Based on these spectra, the equivalent widths (EWs) of the lines of interest 
were measured by using the software SPSHOW (in the SPTOOL\footnote{
$\langle$http://optik2.mtk.nao.ac.jp/\~{ }takeda/sptool/$\rangle$} 
package developed by Y. Takeda) with the Gaussian fitting method or 
the direct-integration method depending on the cases. We did not
use the S~{\sc i} 9228 line for the three F stars, because it is 
blended with the strong Paschen line (H~{\sc i} P$_{9}$) and less 
reliable (cf. figure 3).
Regarding the equivalent width data of the Sun, those of S~{\sc i} 
8693/8694 and 9228/9237 lines were taken from Takada-Hidai et al.'s 
(2005) table 3 (note that the S~{\sc i} 9212 line could not be 
measured because of being heavily blended with a telluric water
vapor line), while the others were newly measured from Kurucz et al.'s (1984) 
solar flux spectrum atlas. The finally resulting EW data used in our 
analysis are presented in table 4.

\subsection{Abundance Results}

The atmospheric parameters of $\alpha$ Per ($T_{\rm eff} = 6250$~K, 
$\log g = 0.90$, $[X]=[{\rm Fe/H}]=0.0$, $\xi$ = 4.5~km~s$^{-1}$)
and Polaris ($T_{\rm eff} = 6000$~K, $\log g = 1.50$, $[X]=[{\rm Fe/H}]=0.0$, 
$\xi$ = 5.0~km~s$^{-1}$) were taken from Takeda and 
Takada-Hidai (1994), while those of Procyon ($T_{\rm eff} = 6600$~K, 
$\log g = 4.00$, $[X]=[{\rm Fe/H}]=0.0$, $\xi$ = 2.0~km~s$^{-1}$)
are the rounded values of the original results derived by Takeda et al. (2005).
Regarding the Sun, we assumed ($T_{\rm eff} = 5780$~K, 
$\log g = 4.44$, $[X]=[{\rm Fe/H}]=0.0$, $\xi$ = 1.0~km~s$^{-1}$).

Again, by using the modified WIDTH9 program as in subsubsection 2.1.3, 
the abundances of S and Zn were derived from the EW data and the model 
atmosphere for each star, which was constructed from Kurucz's (1993) 
ATLAS9 model atmospheres grid by interpolating in terms of $T_{\rm eff}$, 
$\log g$, and [Fe/H]. The resulting NLTE/LTE abundances and the 
corresponding NLTE corrections are given in table 4, where roughly 
estimated values of the mean line-formation depth are also presented.

We should keep in mind that a too-rigorous quantitative discussion is not
very meaningful, especially for $\alpha$ Per and Polaris, because abundance 
determinations of supergiants involve considerable difficulties 
(large uncertainties in establishing $T_{\rm eff}$ and $\log g$, 
depth-dependence of $\xi$, etc.; cf. Takeda, Takada-Hidai 1994). 
Yet, we can recognize that the large/evident discrepancies between 
the LTE abundances of S~{\sc i} 8693/8694 and 9212/9237 ($\ltsim 1.1$ dex 
for $\alpha$ Per,  $\ltsim 0.7$ dex for Polaris, and $\ltsim 0.3$ dex
for Procyon) have been considerably reduced in the case of the NLTE
abundances ($\sim 0.2$ dex for $\alpha$ Per, $\sim 0.1$ dex 
for Polaris\footnote{As seen from the resulting NLTE abundances,
S abundance of Polaris appears to be subsolar by 0.2--0.3 dex,
though we had better be conservative about this, since it might stem 
from errors (e.g., too high $T_{\rm eff}$) in atmospheric parameters.}, 
and $\sim 0.1$ dex for Procyon).\footnote{
Regarding the Sun, the extents of the non-LTE corrections are so small
that they can not be used to judge the adequacy of our non-LTE 
calculations.}
This is surely an encouraging result, which lends support for 
the practical reliability of our non-LTE calculations
(we may state, for example, that $h=0$ is not a bad choice for 
the H~{\sc i} collision rates of neutral sulfur atoms).

In view of the fact that the S~{\sc i} 9212/9228/9237 lines of
multiplet 1 suffer appreciably large non-LTE corrections,
we show in figure 4 how the theoretical non-LTE and LTE profiles 
for the representative S~{\sc i} 9237.54 line (computed with 
the $A^{\rm NLTE}$ value given in table 4) differ from each other, 
and how they match the observed line profile for each star, 
in order to demonstrate the importance of the non-LTE effect.
We can see from this figure that the non-LTE line profiles
satisfactorily reproduce the observations.

Regarding Zn, the non-LTE effect on the zinc abundance determinations
from Zn~{\sc i} lines is insignificant ($\ltsim 0.1$ dex) even in 
the low-gravity F supergiants (as expected), and thus only from 
the present results alone we cannot comment much on the validity of 
our non-LTE calculations on zinc, though we can argue that our 
non-LTE abundances (being practically the same as the LTE abundances) 
do not yield any appreciable inconsistency.

Some remarks on the damping parameters may be due here.
The slight differences between the LTE solar S abundances
of the present study and those of Takada-Hidai et al. (2005)
are mostly due to the differences in the adopted damping parameters.
Namely, in contrast to the present treatment, they used the classical 
formula for the radiation damping, and applied the correction of 
$\Delta\log C_{6} = +0.99$ to the $C_{6}$ value computed from 
the classical Uns\"{o}ld's (1955) formula for the van der Waals effect 
damping, which is equivalent to multiplying $\Gamma_{\rm vdw}^{\rm classical}$ 
by a factor of 2.5 (a frequently assumed enhancement factor).
As a matter of fact, we also investigated how the non-LTE abundances
would change by multiplying the van der Waals damping width 
(for which we assumed the classical treatment, essentially equivalent 
to Uns\"{o}ld's formula, for all lines; cf. table 1) by a factor of 2.5, 
as also shown in table 4. As expected, appreciable negative variations
(amounting to 0.1--0.2 dex) are seen for the high-gravity stars of
the Sun and Procyon. However, since this increase in $\Gamma_{\rm vdw}$ 
deteriorates the consistency between the $A_{\rm NLTE}$ values derived 
from S~{\sc i} 8693/8694 and 9212/9228/9237, we are reluctant to apply
such a correction, considering that the classical treatment 
(as we adopted) is still preferable at least for the lines in question. 
Similarly, almost the same argument holds for the Zn~{\sc i} lines,
as implied from the consistency between $A_{\rm NLTE}$(4722/4810)
and $A_{\rm NLTE}$(6362).

\section{Sulfur Abundances in Metal-Poor Stars}

\subsection{Reanalysis of Literature Data}

We are now ready to study the [S/Fe] vs. [Fe/H] relation
of metal-poor stars by reanalyzing the published equivalent-width
data of the S~{\sc i} 9212/9228/9237 and 8693/8694 lines 
while applying non-LTE corrections based on our calculations.
For this purpose, we invoked the following papers published so far:
Clegg, Lambert, and Tomkin (1981), Fran\c{c}ois (1987, 1988), Israelian and Rebolo (2001),
Takada-Hidai et al. (2002, 2005), Chen et al. (2002), Ryde and Lambert (2004),
and Nissen et al. (2004).
Although our literature survey is not complete, we consider that we have 
picked up most of the important studies, in which the observational data
are explicitly presented.

We adopted the same $T_{\rm eff}$, $\log g$, [Fe/H], and $\xi$ values
as those used in the literature, from which the data of the equivalent widths
were taken. The Kurucz's (1993) grid of ATLAS9 model atmospheres 
and depth-dependent non-LTE departure coefficients were interpolated 
with respect to $T_{\rm eff}$, $\log g$, and [Fe/H] of each star.
Then, as in subsection 3.3, the modified WIDTH9 program was invoked for 
determining the non-LTE abundance ($A_{\rm S}^{\rm NLTE}$) 
while using the line data given in table 1.
Finally, the [S/Fe] ratio was obtained as
\begin{equation}
[{\rm S}/{\rm Fe}] \equiv 
( A_{\rm S}^{\rm NLTE} - 7.21) - [{\rm Fe}/{\rm H}], 
\end{equation}
where 7.21 is the solar sulfur abundance (in the usual scale of 
$\log\epsilon_{\rm H} = 12$) taken from Anders and Grevesse (1989).

In deriving the final [S/Fe] values to be examined, we treated each of 
the [S/Fe] values derived from the S~{\sc i} 9212/9228/9237 lines 
(multiplet 1), S~{\sc i} 8693/8694 lines (multiplet 6), 
S~{\sc i} 6757 line (multiplet 8), and S~{\sc i} 6046/6052 lines 
(multiplet 10), separately, which we hereinafter referred to as
[S/Fe]$_{92}$,  [S/Fe]$_{86}$, [S/Fe]$_{67}$, and [S/Fe]$_{60}$, respectively.
In the case that equivalent-width data are available for more than one S~{\sc i} 
line belonging to the same multiplet, we calculated [S/Fe] 
for each line and adopted their simple mean.\footnote{Regarding 
multiplet 1 (9212/9228/9237), we did not use the S~{\sc i} 9228 line 
in deriving [S/Fe]$_{92}$, because of its less reliability due to 
the blending with H~{\sc i} P$_{9}$.}

The finally resulting [S/Fe] vs. [Fe/H] relation and the 
metallicity dependence of the non-LTE correction are depicted 
in figures 5a and b, respectively.
Also, the details of these analyses (the data of the used equivalent 
widths and the adopted parameter values, the resulting 
non-LTE abundances or [S/Fe] values with the non-LTE corrections, 
given for each line/multiplet and for each star) are given in electronic
table E3 (cf. the footnote in the first page).

In figure 5a, we can see an interesting trend 
concerning the behavior of [S/Fe] and the importance of 
the non-LTE effect in metal-poor stars:\\
--- First, regarding disk stars ($-1 \ltsim$~[Fe/H]~$\ltsim 0$),
the [S/Fe] values of different multiplets almost agree with each other
(though with a rather large dispersion), showing a gradually increasing 
trend with a decrease of metallicity from [S/Fe] $\sim 0$ (at [Fe/H] 
$\sim 0$) to [S/Fe] $\sim$ 0.3 (at [Fe/H] $\sim -1$).\\
--- However, as the metallicity is further lowered, the discrepancy between
[S/Fe]$_{92}$ and [S/Fe]$_{86}$ becomes progressively large,
in the sense that [S/Fe]$_{86}$ continues to rise up to $\sim +0.8$
(at [Fe/H] $\sim -2.5$) while [S/Fe]$_{92}$ shows a sign of leveling-off
at 0.2--0.3 (or even a sign of downward bending at [Fe/H] $\ltsim -2$).

\subsection{9212/9228/9237 vs. 8693/8694 Discrepancy}

Given the existence of such an apparent discrepancy, we have to 
contend with a new question; i.e., which trend better represents 
the truth, [S/Fe]$_{92}$ or [S/Fe]$_{86}$? 

Let us recall here that this difficulty is neither new nor 
restricted to the present analysis including the non-LTE effect.
While our non-LTE correction on [S/Fe]$_{92}$ (by 0.1--0.3 dex acting 
in the direction of suppressing the S abundance; cf. figure 5b) is 
surely responsible for exaggerating this discrepancy, the non-LTE
effect is not the only cause for this discordance. As a matter of fact,
a quantitative inconsistency between [S/Fe]$_{92}$ and [S/Fe]$_{86}$ 
has already been observed even in pure LTE analyses (see, e.g., figure 4 of 
Ryde, Lambert 2004; figure 3 of Takada-Hidai et al. 2005).
As mentioned in section 1, however, the recent S abundance studies 
(e.g., Nissen et al. 2004; Ryde, Lambert 2004) tend to preferably 
invoke [S/Fe]$_{92}$ (on the assumption of LTE) based on the 9212/9228/9237 
lines because of their easy detectability even at very low metallicity, 
while an implicit tendency of neglecting [S/Fe]$_{86}$ appears to exist, 
presumably because it is based on difficult and less reliable 
measurements of weak 8693/8694 lines especially in very metal-poor stars. 
However, is this really a reasonable attitude?
Since the situation is rather complicated, let us sort out our thoughts,
while reconsidering the validity of the so-far adopted assumption.

First of all, it is worth pointing out that any of the recent arguments 
suggesting the validity of LTE for the S~{\sc i} 9212/9228/9237 
lines do not appear to be convincing and should be viewed with caution:\\
--- Nissen et al. (2004) concluded from their analysis
on those stars where both multiplet lines are measurable that the 
non-LTE effect should be insignificant because their LTE abundances 
derived from 8693/8694 and 9212/9237 lines turned out to be in good 
agreement with each other (mean difference is 0.03 dex and the standard 
deviation is 0.08 dex). Actually, we almost confirmed this consequence by
our reanalysis of their data.\footnote{For those 19 stars (out of their 
34 stars) where both multiplet lines available, we obtained
$\langle (A_{86}-A_{92})_{\rm LTE} \rangle = -0.05 (\pm 0.09)$ and
$\langle (A_{86}-A_{92})_{\rm NLTE} \rangle = +0.05 (\pm 0.08)$.}
However, the stars they used for this check were all in the
metallicity range of $-1.8 \ltsim$ [Fe/H] $\ltsim -0.7$, and
did not include any ``very'' metal-deficient stars at [Fe/H] $\ltsim -2$.
This must be a drawback of their argument; i.e., the result obtained 
at $-2 \ltsim$ [Fe/H] should not be simply stretched to the region of 
$-3 \ltsim $ [Fe/H] $\ltsim -2$ without an actual confirmation.
As a matter of fact, as can be seen from figure 5a, the $A_{86}-A_{92}$ 
discrepancy begins to be appreciable at the very such region of 
[Fe/H] $\ltsim -2$. As an example, we can refer to the case of
HD 140283 ([Fe/H] = $-2.42$), for which they did not perform the
$A_{86}$ vs. $A_{92}$ check because its EW(8694) was not measured.
It can be confirmed, however, that $A_{86}$ is appreciably larger than 
$A_{92}$ for this very metal-poor star even in the LTE assumption,
as previously reported in table 3 of Takada-Hidai et al. (2005) as
$(A_{86}-A_{92})_{\rm LTE} = +0.35$ (see also the Appendix for a more
detailed discussion on the $A_{86}$ determination for HD~140283). 
Accordingly, we consider that it is difficult to justify 
Nissen et al.'s (2004) assertion for the practical validity of LTE .\\
--- Ryde and Lambert's (2004) statement that ``NLTE effects
should be small'' because ``neutral sulfur atoms represent that 
main ionization state throughout most of the atmosphere'' is evidently
inadequate, because such a logic can apply only for the case of
departures from LTE in ionization equilibrium (e.g., the frequently
argued case that Fe~{\sc i} lines are affected by non-LTE overionization, 
while Fe~{\sc ii} lines are not). As described in subsection 2.1, since
the non-LTE effect on S~{\sc i} lines is due to the departures from 
LTE excitation, whether or not S~{\sc i} is the dominant ionization 
stage is irrelevant.

We would point out, on the contrary, that there is a good reason 
to suspect the significance of the non-LTE effect, at least for the 
S~{\sc i} 9212/9228/9237 lines of multiplet 1. The transition of this 
multiplet 1 is 4s $^{5}{\rm S}^{\rm o}$ -- 4p $^{5}{\rm P}$,
where the lower $^{5}{\rm S}^{\rm o}$ term (6.52~eV) is ``metastable'' 
(i.e., its radiative connection to the ground triplet term 
3p$^{4}$~$^{3}{\rm P}$ is only via a very weak forbidden transition) 
and the upper 4p $^{5}{\rm P}$ term is just the lower term of the
S~{\sc i} 8693/8694 transition of multiplet 6
(4p $^{5}{\rm P}$ -- 4d $^{5}{\rm D}^{\rm o}$).
It should be stressed here that this situation is quite similar
to the case of neutral oxygen. Namely, the O~{\sc i} 
7771/7774/7775 lines of multiplet 1 
(3s $^{5}{\rm S}^{\rm o}$ -- 3p $^{5}{\rm P}$)
and O~{\sc i} 6156/6158 lines of multiplet 10
(3p $^{5}{\rm P}$ -- 4d $^{5}{\rm D}^{\rm o}$) just correspond to
S~{\sc i} 9212/9228/9237 and S~{\sc i} 8693/8694, respectively.
Then, the formation mechanism of the O~{\sc i} 7771--5 lines
may be informative for understanding the non-LTE effect of 
S~{\sc i} 9212/9228/9237.
Namely, the pseudo two-level-atom nature of the O~{\sc i} 7771--5
line formation (originating from the metastable lower level) may 
approximately also apply to the present case of S~{\sc i} 9212/9228/9237,
which may suffer an appreciable non-LTE effect depending on 
the line-strength, such as the case of the O~{\sc i} triplet lines
at 7771--5 $\rm\AA$ (see, e.g., Takeda 2003).\footnote{
One notable difference between the cases of oxygen and sulfur is that,
while the LTE ionization equilibrium of O~{\sc i}/O~{\sc ii} nearly 
holds due to an efficient charge-exchange reaction leading to
a practically LTE population for the ground level, such a condition
is not expected for the case of S.}
According to this consideration, we had better realize as a starting 
point that LTE may not be a good assumption for S-abundance determinations 
of very metal-poor stars (especially for the triplet lines of multiplet 1 
at 9210--9240 $\rm\AA$), while honestly accepting the discrepancy 
between the LTE abundances of 9212/9228/9237 and 8693/8694 lines.

Yet, this does not solve the currently confronted problem, 
because the non-LTE corrections derived from our calculations act
even in the direction of increasing the discrepancy.
Hence, it is certain that our computed non-LTE corrections 
(for either or both of 9212/9228/9237 and 8693/8694 lines)
are not adequate,\footnote{More precisely, our non-LTE corrections
presented here are not sufficient to reasonably bring the 
abundances from different multiplets into consistency at 
[Fe/H] $\ltsim -2$. While this may imply that our non-LTE calculations 
are simply insufficient/incorrect and inapplicable here, 
an alternative possibility is that some non-classical 
effect (e.g., due to 3D inhomogeneous atmosphere, missing opacity, etc.) 
other than the non-LTE effect may become newly effective at this 
metallicity range.} at least for the purpose of applications to very 
metal-poor stars ([Fe/H] $\ltsim -2$), even if they are successful in 
removing the discordance in the LTE abundances of near-solar metallicity 
F-type stars (cf. section 3). Given such a confusing situation, 
any discussion about the trend (flat? rising?) 
of the [S/Fe] vs. [Fe/H] relation in the very metal-poor regime and 
its comparison with theoretical chemical evolution calculations 
would not be very meaningful (though several representative theoretical
trends are overplotted in figure 5a) until the cause of the problem 
has been clarified. 

\subsection{What Is Wrong and What Should Be Done?}

While we can not decide which of the [S/Fe]$_{92}$ and [S/Fe]$_{86}$ 
is more reliable, some comments (if speculative) may be 
worth presenting here toward a future settlement of this issue:\\
--- It should be kept in mind that our non-LTE calculations turned
out to be successful in reasonably accomplishing an abundance consistency 
for the case of F stars of normal metallicity ($\alpha$ Per, Polaris, 
Procyon). Hence, we feel it rather unlikely that some fatally improper 
modeling of the non-LTE calculation (e.g., a serious mis-choice of $h$, 
lack of important transitions/levels, etc.) is responsible here. 
Instead, we consider that the cause of the flaw in question 
should be such that {\it appearing or becoming evident only at the very low 
metallicity regime}.\\
--- Although we are not qualified to remark on the 3D effect of atmospheric 
inhomogeneity, it seems difficult (at least in a quantitative sense) 
to invoke this effect in order to remove the discrepancy, as long as 
we see the simulation results presented by Nissen et al. (2004).\\
--- There might be a possibility of ``missing opacity'' in our abundance 
calculation program. Namely, if there is some unknown continuum opacity 
(i.e., not included in the WIDTH9 program we adopted) such that being 
less sensitive to the metallicity than the H$^{-}$ opacity, it might 
become significant only in the very metal-poor regime (even if it is 
overwhelmed by H$^{-}$ in the metal-rich case).
Then, the theoretical strengths computed for a given abundance would be 
overestimated, leading to an underestimation of the derived abundances.
Such a problem (if any exists) might be more probable in the 
9210--9240 $\rm\AA$ region, where a close examination (or a photometric 
matching) of the stellar continuum shape is difficult owing to 
crowded telluric water vapor lines, rather than the much better behaved 
8690 $\rm\AA$ region. In any case, such a concern should be checked 
for representative very metal-poor stars by comparing the abundances of 
other elements from this region with those from other regions.\\
--- We still cannot rule out the possibility 
that the problem exists in our non-LTE calculations, 
in the sense that the involved errors/flaws may become 
conspicuous only at the considerably low-metal condition.
Here, we have a suggestion that might be an important touchstone of 
our calculation. Namely, we recommend to observe the 
S~{\sc i} 10455/10456/10459 lines of multiplet 3 for investigating 
the [S/Fe] behavior at [Fe/H] $\ltsim -2$, which are as strong as 
9212/9228/9237 lines (i.e., usable even for very metal-poor stars) 
and favorably located in a wavelength region almost free from 
telluric lines. Since these 10455--9 lines belong to the triplet system 
(4s $^{3}{\rm S}^{\rm o}$ -- 4p $^{3}{\rm P}$)
unlike the quintet-system lines of 9212/9228/9237 and 8693/8694,
it may be a validity check of our non-LTE modeling to compare the 
resulting abundances from different systems with each other.\\
--- Though we would refrain from entering the ``flat vs. rising''
controversy on the [S/Fe] trend, it should be kept in mind that
our understanding of the formation S~{\sc i} 9212/9228/9237 lines 
is still incomplete, which thus should not be trusted too much.
Meanwhile, we should not forget the merit of the weaker and deep-forming
8693/8694 lines (as far as measurements can be accurately done), 
to which the classical line-formation theory may be applied more safely. 
In this sense, the ``rising'' tendency suggested from [S/Fe]$_{86}$ 
(Israelian, Rebolo 2001; Takada-Hidai et al. 2002) might deserve 
more attention.\\
--- In our opinion, what we need is to carry out much more ``reliable'' 
observations of weak S~{\sc i} 8693/8694 lines for as many very 
metal-poor stars ([Fe/H] $\ltsim -2$) as possible, since such data 
are evidently insufficient. Admittedly, the difficulty of 
such observations would progressively increase as we go into 
such a low-metal region, and spectra of very-high S/N would be required. 
It should be here mentioned that very metal-deficient halo stars
of comparatively higher $T_{\rm eff}$ are suitable for this purpose,
since the strengths of such high-excitation S~{\sc i} lines increase 
with $T_{\rm eff}$. Hence, as a possibility, it may be promising to 
pay attention to F-type horizontal-branch stars in very metal-poor 
globular clusters (e.g., M92; [Fe/H] = $-2.3$).  

\section{Behavior of Zinc in Disk/Halo Stars}

Similarly to the case of sulfur, we also carried out an extensive non-LTE
reanalysis of the published equivalent-width of Zn~{\sc i} 4722/4810
and 6362 lines taken from the following papers:\footnote{We remark that
there are other important recent papers apart from these eight 
concerning Zn abundance analyses for a number of stars, such as 
Primas et al. (2000), Mishenina et al. (2002), Reddy et al. (2003), 
Bensby, Feltzing, and Lundstr\"{o}m (2003), Allende Prieto et al. (2004), and 
Ecuvillon et al. (2004). Unfortunately, since these authors presented
only the abundance results and did not publish observed EW values for
each star, we could not include their data in our non-LTE reanalyses.}
Sneden and Crocker (1988), Sneden, Gratton, and Crocker (1991), 
Beveridge and Sneden (1994), Prochaska et al. (2000), Nissen et al. (2004), 
Cayrel et al. (2004), Honda et al. (2004), and Chen, Nissen, and Zhao (2004).
In a similar way as described in subsection 4.1, non-LTE Zn 
abundances ($A_{\rm Zn}^{\rm NLTE}$) were determined
from these EW data of 4722/4810/6362\footnote{ 
In deriving the abundance from the Zn~{\sc i} 6362 line, we did not
take into account the broad wing of the Ca~{\sc i} autoionization line,
since it is practically negligible (cf. Chen et al. 2004).} lines
along with the model atmospheres corresponding to
the atmospheric parameters taken from the same papers of the EW source.
The [Zn/Fe] ratio was derived as
\begin{equation}
[{\rm Zn}/{\rm Fe}] \equiv 
( A_{\rm Zn}^{\rm NLTE} - 4.60) - [{\rm Fe}/{\rm H}],
\end{equation}
where 4.60 is the solar zinc abundance (Anders, Grevesse 1989).
Treating the Zn~{\sc i} 4722/4810 lines (multiplet 2) and the 6362 line 
(multiplet 6) separately, we derived [Zn/Fe]$_{4722/4810}$ and 
[Zn/Fe]$_{6362}$. In the case where both of the 4722 and 4810 lines are 
available, we adopted a simple average of the two to obtain 
[Zn/Fe]$_{4722/4810}$. 
The finally resulting [Zn/Fe] vs. [Fe/H] relation and the 
metallicity-dependence of the non-LTE correction are depicted 
in figures 6a and b, respectively.
As in the case of S, the details of these Zn reanalyses are given 
in electronic table E4 (cf. the footnote in the first page).

By inspecting figures 6a and b, we can see the following characteristics:\\
--- Generally speaking, the non-LTE corrections in zinc abundance 
determinations from Zn~{\sc i} 4722/4810 and 6362 lines are comparatively
insignificant ($\ltsim 0.1$ dex in most cases). Especially, LTE is
a practically valid approximation for the Zn~{\sc i} 6362 line, which is 
appropriate for investigating the Zn abundances of disk stars 
as recently used by Chen, Nissen, and Zhao (2004). 
The corrections are mostly positive (i.e., the non-LTE effect tends to act
in the direction of weakening the line) except for some cases of the 
4722/4810 lines, and have a slightly increasing trend with a lowering
of the metallicity, which gives a mild influence on the [Zn/Fe] vs. [Fe/H] 
relation as described below.\\
--- The traditional belief that ``Zn almost scales with Fe over 
a wide range of metallicity (i.e., [Zn/Fe] $\sim 0$)'' does not 
adequately represent the truth from a strict point of view.
Namely, the [Zn/Fe] values for disk stars gradually increase from 
[Zn/Fe] $\sim 0$ (at [Fe/H] $\sim 0$) to [Zn/Fe] $\sim$ 0.2 
(at [Fe/H] $\sim -1$). While it appears that a kind of weak discontinuity 
exists at [Fe/H] $\sim -1$, [Zn/Fe] exhibits a pseudo-plateau at 
$\sim 0.2$ (or a slightly increasing trend with a very gentle slope)
over the region of $-2 \ltsim$ [Fe/H] $\ltsim -1$. Then, 
below [Fe/H] $\sim -2$, [Zn/Fe] begins to show a manifest rise 
toward an extremely low metallicity regime, which continues
down to [Fe/H] $\sim -4$ where [Zn/Fe] attains even a value of $\sim +1$.
In short, we have simply confirmed the recently reported results, such as
the slight increase of [Zn/Fe] for disk stars with a decrease of [Fe/H]
(Reddy et al. 2003; Chen et al. 2004) and the clear evidence of 
markedly rising [Zn/Fe] at [Fe/H] $\ltsim -2.5$ (Primas et al. 2000; 
Cayrel et al. 2004), while these trends have become even more
pronounced in our reanalyses because of the (positive) non-LTE corrections
gradually increasing with a decrease of metallicity.\\
--- It appears that none of the theoretical predictions from representative 
Galactic chemical evolution calculations (such those overplotted in 
figure 6a) adequately explain such a trend of [Zn/Fe] with a run of [Fe/H],
though (somewhat interestingly) only the calculation by Goswami and 
Prantzos (2000) under the assumption of ``constant yield fixed at the 
solar metallicity'' (i.e., labeled ``GP(const)''; an unrealistic case 
only for an illustration purpose to show the significance of 
the metallicity-dependence of the yield) appears to roughly show 
a kind of similar tendency (though only qualitatively, 
not quantitatively). New theoretical calculations that can adequately 
reproduce such an observationally established [Zn/Fe] vs. [Fe/H] relation
are awaited.

\section{Concluding Summary}

We carried out statistical-equilibrium calculations on neutral sulfur
and zinc for an extensive grid of models over a wide range of parameters
($T_{\rm eff}$ from 4500~K to 7000~K, $\log g$ from 1.0 to 5.0, and
[Fe/H] from $-4.0$ to +0.5), in order to study how the non-LTE effect
is important in deriving the photospheric S and Zn abundances of 
F, G, and K stars from various S~{\sc i} and Zn~{\sc i} lines,
where a particular emphasis was placed on S~{\sc i} 8693/8694 lines
(multiplet 6), S~{\sc i} 9212/9228/9237 lines (multiplet 1), 
Zn~{\sc i} 4722/4810 lines (multiplet 2), and Zn~{\sc i} 6362 line
(multiplet 6). We then constructed an extensive grid of non-LTE 
abundance corrections, which should be applied to the conventionally 
derived LTE abundance to obtain the corresponding non-LTE abundance.

Roughly speaking, the non-LTE effect on S~{\sc i} lines acts
in the direction of strengthening the lines (i.e., negative
non-LTE correction) owing to the overpopulation of the lower level
coupled with the dilution of the line source function. 
On the other hand, the line source function for the Zn~{\sc i} lines
tends to be superthermal ($S_{\rm L} > B$), while the lower level shows
either non-LTE overpopulation or underpopulation depending on 
the cases; as a result, Zn~{\sc i} lines tend to be weakened
(i.e., positive correction) by the non-LTE effect in many cases, 
though the opposite can occur in some cases.
Quantitatively, the extents of the non-LTE corrections for the S~{\sc i} 
9212/9228/9237 lines are considerably larger and more important than 
those for the S~{\sc i} 8693/8694 lines, while the non-LTE effect for 
the Zn~{\sc i} lines is generally of minor importance.

Taking account of the fact that the non-LTE effect tends to become
larger with an increase/lowering of $T_{\rm eff}$/$\log g$,
we performed abundance analyses of sulfur and zinc for 
representative F supergiants/subgiant ($\alpha$ Per, Polaris, Procyon)
along with the Sun, in order to check the validity of our 
non-LTE calculations by examining whether a consistency can be
achieved between the abundances derived from different lines.
For this purpose, we used the high-dispersion echelle spectra
obtained with the GAOES spectrograph at Gunma Astronomical
Observatory. It was confirmed that the large discrepancies
seen in the LTE S abundances of these F stars could be successfully
removed by our non-LTE corrections, while the non-LTE corrections
for the Zn~{\sc i} lines were too small to be useful for such a check. 

Finally, extensive non-LTE reanalyses of published equivalent-width 
data of the S~{\sc i} and Zn~{\sc i} lines were carried out, in order to 
investigate the behavior of [S/Fe] or [Zn/Fe] with a change of 
[Fe/H] in Galactic disk/halo stars.
The following conclusions were reached from this restudy:\\
--- We encountered a serious difficulty in the [S/Fe] vs. [Fe/H] 
relation at the very metal-poor region of [Fe/H] $\ltsim -2$,
in the sense that the [S/Fe] values derived from S~{\sc i} 8693/8694
tend to rise progressively with a decrease of metallicity
while those from S~{\sc i} 9212/9228/9237 show a flat plateau
(or a sign of slight downward bending); the discordance
amounts up to $\sim 0.5$ dex. Though such a trend has actually been
reported even in the framework of LTE, this discrepancy has been
exaggerated due to the application of non-LTE corrections. 
At present, little can be said about which represents the truth. 
Yet, we have a rather conservative feeling that the solution based on
the weaker and deep-forming 8693/8694 lines (ever-rising [S/Fe]) 
might deserve more attention (in contrast to the arguments of
recent studies), since our understanding of 9212/9228/9237 lines 
is still incomplete, with which unclarified problems might be 
involved (e.g., flaws in non-LTE calculations, 3D effect, 
missing opacity, etc.).\\
--- Inspecting the [Zn/Fe] vs. [Fe/H] relation resulting from our
reanalysis, we almost confirmed the recently reported tendency,
such as the gradual increase of [Zn/Fe] from [Zn/Fe] $\sim 0$ 
(at [Fe/H] $\sim 0$) to [Zn/Fe] $\sim$ 0.2 (at [Fe/H] $\sim -1$), 
nearly constant  [Zn/Fe] (i.e., without particular trend) over the 
region of $-2 \ltsim$ [Fe/H] $\ltsim -1$, and a beginning of rise at 
[Fe/H] $\ltsim -2$ continuing toward an extremely low-metallicity regime.
Actually, these trends appear even more exaggerated in our 
reanalysis results, because the (positive) non-LTE corrections acting on 
the Zn abundances are metallicity-dependent (i.e., progressively
increasing with a decrease of metallicity).

\appendix

\section*{Sulfur 8694.63 Line in the Spectrum of HD~140283}

Since the S abundance determination of very metal-poor stars 
([Fe/H] $< -2$) based on the S~{\sc i} 8694 line is very difficult 
because of its considerable weakness, the rare case of HD~140283
is especially important, for which one may manage to detect and measure 
this line as recently carried out by Takada-Hidai et al. (2005). 
Since only the figure in a magnified scale that they presented for 
demonstrating the rather delicate S~{\sc i} 8694 detection (cf. 
their figure 2d) is not necessarily sufficient for the reader to 
judge its reliability, we show here some supplementary and more 
informative figures.

As Takada-Hidai et al. (2005) did, we invoked the ESO/UVES spectrum
of HD~140283 in the published high-dispersion stellar spectral library, 
``A Library of High-Resolution Spectra of Stars across the 
Hertzsprung-Russell Diagram'' (Bagnulo et al. 2003). Regarding the
calculation of theoretical spectra to be compared with observations,
we adopted the atmospheric parameters of $T_{\rm eff}$ = 5960~K, 
$\log g$ = 3.69, $v_{\rm t}$ = 1.5~km~s$^{-1}$, and [Fe/H] = $-2.42$,
which were taken from Nissen et al. (2004). Since the non-LTE effect
is practically negligible for such a weak S~{\sc i} 8694 line (cf. 
electronic table E3), we assumed LTE in the spectrum synthesis.
A comparison of the observed spectrum with three computed spectra
corresponding to [S/Fe] = 0.0, +0.5, and +1.0 is displayed in figures 
7a (wide view) and b (magnified view). The theoretical spectra 
are convolved with a Gaussian broadening function, which was so chosen 
as to accomplish the best fit for the conspicuously seen Fe~{\sc i} 
8688.62 line (cf. figure 7a). 

Inspecting these figures, we can see that a weak (but recognizable) dip 
with a depth of $\sim$~1\% surely exists at the position of 
S~{\sc i} 8694.63. Admittedly, we cannot rule out a possibility 
that this is nothing but a fluctuation of fringe patterns. 
However, since the S/N ratio of this spectrum is estimated to be 
$\sim 500$ ($\sigma \sim$ 0.2\%) from the line-free 
8690--8692 $\rm\AA$ region, the possibility of such a large 
fluctuation (amounting to $\sim 5\sigma$) is not considered to be 
very likely. Then, on the standpoint that this identification is
real, we may state that [S/Fe] should be near to $\sim +1$, since
this line would not be visible if [S/Fe] $\ltsim +0.5$ (figure 7b).

\clearpage
\setcounter{table}{0}
\begin{table}[h]
\scriptsize
\caption{Atomic data of the relevant S~{\sc i} and Zn~{\sc i} lines.}
\begin{center}
\begin{tabular}
{crccrccccc}\hline \hline
Species & Line & RMT & Multiplet & $\lambda$ ($\rm\AA)$ & $\chi_{\rm low}$ (eV) & $\log gf$ & Gammar & Gammas & Gammaw \\ \hline

S~{\sc i} & 9212 & 1 &  4s $^{5}{\rm S}^{\rm o}_{2}$ -- 4p $^{5}{\rm P}_{3}$ & 9212.863 & 6.524 & +0.420 & 7.47 & $-5.24$ & ($-7.60$) \\
S~{\sc i} &  9228 & 1 &  4s $^{5}{\rm S}^{\rm o}_{2}$ -- 4p $^{5}{\rm P}_{2}$ & 9228.093 & 6.524 & +0.260 & 7.46 & $-5.24$ & ($-7.60$) \\
S~{\sc i} &  9237 & 1 &  4s $^{5}{\rm S}^{\rm o}_{2}$ -- 4p $^{5}{\rm P}_{1}$ & 9237.538 & 6.524 & +0.040 & 7.46 & $-5.24$ & ($-7.60$) \\
S~{\sc i} & 10455 & 3 &  4s $^{3}{\rm S}^{\rm o}_{1}$ -- 4p $^{3}{\rm P}_{2}$ & 10455.449 & 6.860 & +0.260 & 8.86 & $-5.21$ & ($-7.57$) \\
S~{\sc i} &  10456 & 3 &  4s $^{3}{\rm S}^{\rm o}_{1}$ -- 4p $^{3}{\rm P}_{0}$ & 10456.757 & 6.860 & $-0.430$ & 8.86 & $-5.21$ & ($-7.57$) \\
S~{\sc i} &  10459 & 3 &  4s $^{3}{\rm S}^{\rm o}_{1}$ -- 4p $^{3}{\rm P}_{1}$ & 10459.406 & 6.860 & +0.040 & 8.86 & $-5.21$ & ($-7.57$) \\
S~{\sc i} &  8693 & 6 &   4p $^{5}{\rm P}_{3}$ -- 4d $^{5}{\rm D}_{3}^{\rm o}$ & 8693.931 & 7.870 & $-0.510$ & 7.62 & $-4.41$ & ($-7.30$) \\
S~{\sc i} &  8694 & 6 &   4p $^{5}{\rm P}_{3}$ -- 4d $^{5}{\rm D}_{4}^{\rm o}$ & 8694.626 & 7.870 & +0.080 & 7.62 & $-4.41$ & ($-7.30$) \\
S~{\sc i} &  6757 & 8 &  4p $^{5}{\rm P}_{3}$ -- 5d $^{5}{\rm D}_{2}^{\rm o}$ & 6756.851 & 7.870 & $-1.760$ & 7.59 & $-3.86$ & ($-7.13$) \\
        &       & 8 &  4p $^{5}{\rm P}_{3}$ -- 5d $^{5}{\rm D}_{3}^{\rm o}$ & 6757.007 & 7.870 & $-0.900$ & 7.59 & $-3.86$ & ($-7.13$) \\
        &       & 8 &  4p $^{5}{\rm P}_{3}$ -- 5d $^{5}{\rm D}_{4}^{\rm o}$ & 6757.171 & 7.870 & $-0.310$ & 7.59 & $-3.86$ & ($-7.13$) \\
S~{\sc i} &  6046 & 10 &  4p $^{5}{\rm P}_{2}$ -- 6d $^{5}{\rm D}_{1}^{\rm o}$ & 6045.954 & 7.867 & $-1.820$ & (7.78) & ($-4.28$) & ($-7.00$) \\
         &      & 10 &  4p $^{5}{\rm P}_{2}$ -- 6d $^{5}{\rm D}_{2}^{\rm o}$ & 6045.991 & 7.867 & $-1.240$ & (7.78) & ($-4.28$) & ($-7.00$) \\
         &      & 10 &  4p $^{5}{\rm P}_{2}$ -- 6d $^{5}{\rm D}_{3}^{\rm o}$ & 6046.027 & 7.867 & $-1.030$ & (7.78) & ($-4.28$) & ($-7.00$) \\
S~{\sc i} &  6052 & 10 &  4p $^{5}{\rm P}_{3}$ -- 6d $^{5}{\rm D}_{3}^{\rm o}$ & 6052.583 & 7.870 & $-1.330$ & (7.78) & ($-4.28$) & ($-7.00$) \\
          &     & 10 &  4p $^{5}{\rm P}_{3}$ -- 6d $^{5}{\rm D}_{4}^{\rm o}$ & 6052.674 & 7.870 & $-0.740$ & (7.78) & ($-4.28$) & ($-7.00$) \\ \hline
Zn~{\sc i} &  4722 & 2 &  4p $^{3}{\rm P}_{1}^{\rm o}$ -- 5s $^{3}{\rm S}_{1}$ & 4722.153 & 4.030 & $-0.338$ & (8.00) & ($-6.26$) & ($-7.63$) \\
Zn~{\sc i} &  4810 & 2 &  4p $^{3}{\rm P}_{2}^{\rm o}$ -- 5s $^{3}{\rm S}_{1}$ & 4810.528 & 4.078 & $-0.137$ & (7.98) & ($-6.26$) & ($-7.63$) \\
Zn~{\sc i} &  6362 & 6 &  4p $^{1}{\rm P}_{1}^{\rm o}$ -- 4d $^{1}{\rm D}_{2}$ & 6362.338 & 5.796 & $+0.150$ & (7.74) & ($-5.71$) & ($-7.45$) \\
\hline
\end{tabular}
\end{center}
All data are were taken from Kurucz and Bell's (1995) compilation
as far as available. RMT is the multiplet number given by the 
Revised Multiplet Table (Moore 1959).
Gammar is the radiation damping constant, $\log\gamma_{\rm rad}$.
Gammas is the Stark damping width per electron density
at $10^{4}$ K, $\log(\gamma_{\rm e}/N_{\rm e})$.
Gammaw is the van der Waals damping width per hydrogen density
at $10^{4}$ K, $\log(\gamma_{\rm w}/N_{\rm H})$. 
Note that the values in parentheses are the default damping parameters 
computed within the Kurucz's WIDTH program (cf. Leusin, Topil'skaya 1987),
because of being unavailable in Kurucz and Bell (1995).
The meanings of other columns are self-explanatory.

\end{table}

\clearpage
\setcounter{table}{1}
\begin{table}[h]
\tiny
\caption{Dependence of the non-LTE effect of S~{\sc i} 8694 and 9212 lines
on the H~{\sc i} collision and photoionization cross section.} 
\begin{center}
\begin{tabular}
{ccrcccrr@{ }r@{ }r@{ }r@{ }rr@{ }r@{ }r@{ }r@{ }rr@{}r}\hline \hline
$T_{\rm eff}$  & $\log g$ & [Fe/H] & $\xi$ & $A_{\rm S}$ & Line &
$W^{\rm LTE}$ & $W_{+1}$ & $W_{0}$ & $W_{-1}$ & $W_{-2}$ & $W_{-3}$ &
$\Delta_{+1}$ & $\Delta_{0}$ & $\Delta_{-1}$ & $\Delta_{-2}$ & $\Delta_{-3}$
& $\delta(\Delta)_{-}$ & $\delta(\Delta)_{+}$ \\
\hline
4500 & 4.0 & $-$1.0 &  2.0 &  6.71 &  8694.63 &    1.6 &    1.6 & {\bf   1.7}  &   1.7 &   1.9 &   2.1 &   0.00 & {\bf  0.00}  & $-$0.01 & $-$0.06 & $-$0.11  &   0.00 &   0.00 \\
4500 & 4.0 & $-$2.0 &  2.0 &  5.71 &  8694.63 &    0.2 &    0.2 & {\bf   0.2}  &   0.2 &   0.2 &   0.3 &   0.00 & {\bf  0.00}  &  0.00 & $-$0.08 & $-$0.22  &   0.00 &   0.00 \\
4500 & 4.0 & $-$3.0 &  2.0 &  4.71 &  8694.63 &    0.0 &    0.0 & {\bf   0.0}  &   0.0 &   0.0 &   0.1 &   0.00 & {\bf  0.00}  &  0.00 & $-$0.12 & $-$0.40  &   0.00 &   0.00 \\
4500 & 4.0 & $-$4.0 &  2.0 &  3.71 &  8694.63 &    0.0 &    0.0 & {\bf   0.0}  &   0.0 &   0.0 &   0.0 &   0.00 & {\bf  0.00}  & $-$0.02 & $-$0.23 & $-$0.50  &   0.00 &   0.00 \\
5500 & 2.0 &  0.0 &  2.0 &  7.21 &  8694.63 &   63.1 &   66.1 & {\bf  75.9}  &  87.1 &  91.2 &  91.2 &  $-$0.05 & {\bf $-$0.21}  & $-$0.38 & $-$0.44 & $-$0.45  &  $-$0.01 &  +0.08 \\
5500 & 2.0 & $-$1.0 &  2.0 &  6.71 &  8694.63 &   35.5 &   36.3 & {\bf  41.7}  &  51.3 &  56.2 &  57.5 &  $-$0.02 & {\bf $-$0.13}  & $-$0.29 & $-$0.38 & $-$0.40  &  $-$0.01 &  +0.05 \\
5500 & 2.0 & $-$2.0 &  2.0 &  5.71 &  8694.63 &    5.9 &    5.9 & {\bf   7.1}  &  10.7 &  13.5 &  14.1 &  $-$0.01 & {\bf $-$0.09}  & $-$0.30 & $-$0.43 & $-$0.45  &   0.00 &  +0.03 \\
5500 & 2.0 & $-$3.0 &  2.0 &  4.71 &  8694.63 &    0.6 &    0.6 & {\bf   1.0}  &   2.1 &   2.8 &   2.9 &  $-$0.02 & {\bf $-$0.22}  & $-$0.55 & $-$0.67 & $-$0.68  &   0.00 &  +0.04 \\
5500 & 2.0 & $-$4.0 &  2.0 &  3.71 &  8694.63 &    0.1 &    0.1 & {\bf   0.2}  &   0.3 &   0.4 &   0.4 &  $-$0.10 & {\bf $-$0.47}  & $-$0.75 & $-$0.82 & $-$0.83  &  +0.04 &  +0.01 \\
5500 & 4.0 &  0.0 &  2.0 &  7.21 &  8694.63 &   27.5 &   27.5 & {\bf  28.2}  &  30.9 &  33.1 &  33.9 &   0.00 & {\bf $-$0.01}  & $-$0.07 & $-$0.13 & $-$0.15  &   0.00 &  +0.01 \\
5500 & 4.0 & $-$1.0 &  2.0 &  6.71 &  8694.63 &   12.0 &   12.0 & {\bf  12.3}  &  13.2 &  15.5 &  17.4 &   0.00 & {\bf  0.00}  & $-$0.05 & $-$0.14 & $-$0.19  &   0.00 &   0.00 \\
5500 & 4.0 & $-$2.0 &  2.0 &  5.71 &  8694.63 &    1.3 &    1.3 & {\bf   1.3}  &   1.5 &   2.2 &   3.1 &   0.00 & {\bf  0.00}  & $-$0.05 & $-$0.24 & $-$0.38  &   0.00 &   0.00 \\
5500 & 4.0 & $-$3.0 &  2.0 &  4.71 &  8694.63 &    0.1 &    0.1 & {\bf   0.1}  &   0.2 &   0.5 &   0.7 &   0.00 & {\bf $-$0.01}  & $-$0.17 & $-$0.55 & $-$0.73  &   0.00 &   0.00 \\
5500 & 4.0 & $-$4.0 &  2.0 &  3.71 &  8694.63 &    0.0 &    0.0 & {\bf   0.0}  &   0.0 &   0.1 &   0.1 &  $-$0.01 & {\bf $-$0.09}  & $-$0.45 & $-$0.83 & $-$0.96  &  +0.01 &  $-$0.02 \\
6500 & 2.0 &  0.0 &  2.0 &  7.21 &  8694.63 &   97.7 &  109.7 & {\bf 128.8}  & 134.9 & 138.0 & 138.0 &  $-$0.17 & {\bf $-$0.42}  & $-$0.52 & $-$0.55 & $-$0.55  &  $-$0.04 &  +0.33 \\
6500 & 2.0 & $-$1.0 &  2.0 &  6.71 &  8694.63 &   61.7 &   69.2 & {\bf  81.3}  &  89.1 &  89.1 &  89.1 &  $-$0.12 & {\bf $-$0.31}  & $-$0.42 & $-$0.43 & $-$0.44  &  $-$0.03 &  +0.21 \\
6500 & 2.0 & $-$2.0 &  2.0 &  5.71 &  8694.63 &   12.6 &   14.1 & {\bf  19.5}  &  24.5 &  25.1 &  25.1 &  $-$0.05 & {\bf $-$0.23}  & $-$0.35 & $-$0.37 & $-$0.37  &  $-$0.03 &  +0.15 \\
6500 & 2.0 & $-$3.0 &  2.0 &  4.71 &  8694.63 &    1.2 &    1.7 & {\bf   3.5}  &   4.5 &   4.7 &   4.7 &  $-$0.16 & {\bf $-$0.47}  & $-$0.59 & $-$0.61 & $-$0.61  &  +0.05 &  +0.20 \\
6500 & 2.0 & $-$4.0 &  2.0 &  3.71 &  8694.63 &    0.1 &    0.2 & {\bf   0.5}  &   0.6 &   0.7 &   0.7 &  $-$0.25 & {\bf $-$0.60}  & $-$0.72 & $-$0.74 & $-$0.74  &  +0.14 &  +0.12 \\
6500 & 4.0 &  0.0 &  2.0 &  7.21 &  8694.63 &   63.1 &   63.1 & {\bf  66.1}  &  75.9 &  81.3 &  81.3 &  $-$0.01 & {\bf $-$0.06}  & $-$0.18 & $-$0.24 & $-$0.25  &  $-$0.01 &  +0.03 \\
6500 & 4.0 & $-$1.0 &  2.0 &  6.71 &  8694.63 &   30.2 &   30.2 & {\bf  31.6}  &  36.3 &  40.7 &  41.7 &   0.00 & {\bf $-$0.04}  & $-$0.13 & $-$0.20 & $-$0.21  &   0.00 &  +0.02 \\
6500 & 4.0 & $-$2.0 &  2.0 &  5.71 &  8694.63 &    4.0 &    4.0 & {\bf   4.3}  &   5.6 &   7.1 &   7.4 &   0.00 & {\bf $-$0.02}  & $-$0.15 & $-$0.26 & $-$0.29  &   0.00 &  +0.02 \\
6500 & 4.0 & $-$3.0 &  2.0 &  4.71 &  8694.63 &    0.4 &    0.4 & {\bf   0.6}  &   1.1 &   1.7 &   1.8 &  $-$0.01 & {\bf $-$0.12}  & $-$0.45 & $-$0.62 & $-$0.65  &  +0.01 &  +0.02 \\
6500 & 4.0 & $-$4.0 &  2.0 &  3.71 &  8694.63 &    0.0 &    0.1 & {\bf   0.1}  &   0.2 &   0.3 &   0.3 &  $-$0.06 & {\bf $-$0.29}  & $-$0.66 & $-$0.82 & $-$0.85  &  +0.05 &  $-$0.03 \\
\hline
4500 & 2.0 &  0.0 &  2.0 &  7.21 &  9212.86 &   97.7 &  109.7 & {\bf 123.0}  & 134.9 & 141.2 & 141.2 &  $-$0.14 & {\bf $-$0.32}  & $-$0.46 & $-$0.52 & $-$0.54  &   0.00 &   0.00 \\
4500 & 2.0 & $-$1.0 &  2.0 &  6.71 &  9212.86 &   85.1 &   91.2 & {\bf 102.3}  & 117.5 & 128.8 & 134.9 &  $-$0.08 & {\bf $-$0.25}  & $-$0.41 & $-$0.56 & $-$0.62  &   0.00 &   0.00 \\
4500 & 2.0 & $-$2.0 &  2.0 &  5.71 &  9212.86 &   39.8 &   41.7 & {\bf  45.7}  &  55.0 &  66.1 &  74.1 &  $-$0.02 & {\bf $-$0.11}  & $-$0.25 & $-$0.42 & $-$0.53  &   0.00 &   0.00 \\
4500 & 2.0 & $-$3.0 &  2.0 &  4.71 &  9212.86 &    7.9 &    8.1 & {\bf   9.3}  &  12.6 &  18.2 &  22.4 &  $-$0.01 & {\bf $-$0.08}  & $-$0.23 & $-$0.43 & $-$0.55  &   0.00 &   0.00 \\
4500 & 2.0 & $-$4.0 &  2.0 &  3.71 &  9212.86 &    0.9 &    0.9 & {\bf   1.1}  &   1.8 &   2.6 &   3.2 &  $-$0.02 & {\bf $-$0.11}  & $-$0.34 & $-$0.51 & $-$0.59  &   0.00 &   0.00 \\
4500 & 4.0 &  0.0 &  2.0 &  7.21 &  9212.86 &   55.0 &   55.0 & {\bf  57.5}  &  63.1 &  66.1 &  67.6 &   0.00 & {\bf $-$0.04}  & $-$0.11 & $-$0.17 & $-$0.19  &   0.00 &   0.00 \\
4500 & 4.0 & $-$1.0 &  2.0 &  6.71 &  9212.86 &   45.7 &   45.7 & {\bf  46.8}  &  51.3 &  57.5 &  61.7 &   0.00 & {\bf $-$0.02}  & $-$0.08 & $-$0.16 & $-$0.23  &   0.00 &   0.00 \\
4500 & 4.0 & $-$2.0 &  2.0 &  5.71 &  9212.86 &   12.9 &   12.9 & {\bf  12.9}  &  14.1 &  17.4 &  22.9 &   0.00 & {\bf $-$0.01}  & $-$0.05 & $-$0.16 & $-$0.29  &   0.00 &   0.00 \\
4500 & 4.0 & $-$3.0 &  2.0 &  4.71 &  9212.86 &    1.7 &    1.7 & {\bf   1.7}  &   2.0 &   3.1 &   5.1 &   0.00 & {\bf $-$0.01}  & $-$0.09 & $-$0.27 & $-$0.50  &   0.00 &   0.00 \\
4500 & 4.0 & $-$4.0 &  2.0 &  3.71 &  9212.86 &    0.2 &    0.2 & {\bf   0.2}  &   0.2 &   0.5 &   0.7 &   0.00 & {\bf $-$0.02}  & $-$0.14 & $-$0.42 & $-$0.61  &   0.00 &   0.00 \\
5500 & 2.0 &  0.0 &  2.0 &  7.21 &  9212.86 &  186.2 &  223.9 & {\bf 257.0}  & 281.8 & 288.4 & 295.1 &  $-$0.38 & {\bf $-$0.63}  & $-$0.79 & $-$0.85 & $-$0.85  &  $-$0.01 &  +0.02 \\
5500 & 2.0 & $-$1.0 &  2.0 &  6.71 &  9212.86 &  151.4 &  182.0 & {\bf 213.8}  & 245.5 & 263.0 & 263.0 &  $-$0.40 & {\bf $-$0.74}  & $-$0.98 & $-$1.09 & $-$1.11  &  $-$0.01 &  +0.03 \\
5500 & 2.0 & $-$2.0 &  2.0 &  5.71 &  9212.86 &   81.3 &   95.5 & {\bf 123.0}  & 162.2 & 186.2 & 186.2 &  $-$0.22 & {\bf $-$0.65}  & $-$1.21 & $-$1.48 & $-$1.50  &  $-$0.02 &  +0.05 \\
5500 & 2.0 & $-$3.0 &  2.0 &  4.71 &  9212.86 &   21.9 &   27.5 & {\bf  47.9}  &  81.3 &  93.3 &  95.5 &  $-$0.13 & {\bf $-$0.51}  & $-$1.01 & $-$1.20 & $-$1.23  &  $-$0.01 &  +0.07 \\
5500 & 2.0 & $-$4.0 &  2.0 &  3.71 &  9212.86 &    2.8 &    4.8 & {\bf  12.3}  &  19.5 &  22.4 &  22.9 &  $-$0.24 & {\bf $-$0.69}  & $-$0.94 & $-$1.01 & $-$1.02  &  +0.02 &  +0.04 \\
5500 & 4.0 &  0.0 &  2.0 &  7.21 &  9212.86 &  151.4 &  154.9 & {\bf 169.8}  & 195.0 & 213.8 & 213.8 &  $-$0.04 & {\bf $-$0.16}  & $-$0.33 & $-$0.43 & $-$0.45  &   0.00 &   0.00 \\
5500 & 4.0 & $-$1.0 &  2.0 &  6.71 &  9212.86 &  123.0 &  125.9 & {\bf 138.0}  & 162.2 & 190.6 & 204.2 &  $-$0.03 & {\bf $-$0.12}  & $-$0.32 & $-$0.51 & $-$0.60  &   0.00 &  +0.01 \\
5500 & 4.0 & $-$2.0 &  2.0 &  5.71 &  9212.86 &   40.7 &   41.7 & {\bf  44.7}  &  57.5 &  85.1 & 107.2 &  $-$0.01 & {\bf $-$0.06}  & $-$0.24 & $-$0.55 & $-$0.78  &   0.00 &   0.00 \\
5500 & 4.0 & $-$3.0 &  2.0 &  4.71 &  9212.86 &    5.9 &    6.0 & {\bf   7.1}  &  13.2 &  28.8 &  38.9 &  $-$0.01 & {\bf $-$0.08}  & $-$0.37 & $-$0.78 & $-$0.96  &   0.00 &   0.00 \\
5500 & 4.0 & $-$4.0 &  2.0 &  3.71 &  9212.86 &    0.6 &    0.7 & {\bf   1.0}  &   3.2 &   6.3 &   7.8 &  $-$0.03 & {\bf $-$0.22}  & $-$0.72 & $-$1.03 & $-$1.13  &   0.00 &   0.00 \\
6500 & 2.0 &  0.0 &  2.0 &  7.21 &  9212.86 &  204.2 &  263.0 & {\bf 295.1}  & 309.0 & 309.0 & 309.0 &  $-$0.59 & {\bf $-$0.81}  & $-$0.91 & $-$0.93 & $-$0.94  &  $-$0.02 &  +0.08 \\
6500 & 2.0 & $-$1.0 &  2.0 &  6.71 &  9212.86 &  154.9 &  218.8 & {\bf 257.0}  & 275.4 & 275.4 & 275.4 &  $-$0.80 & {\bf $-$1.13}  & $-$1.26 & $-$1.29 & $-$1.29  &  $-$0.02 &  +0.12 \\
6500 & 2.0 & $-$2.0 &  2.0 &  5.71 &  9212.86 &   91.2 &  134.9 & {\bf 177.8}  & 199.5 & 199.5 & 199.5 &  $-$0.75 & {\bf $-$1.44}  & $-$1.71 & $-$1.74 & $-$1.74  &  $-$0.05 &  +0.28 \\
6500 & 2.0 & $-$3.0 &  2.0 &  4.71 &  9212.86 &   27.5 &   51.3 & {\bf  87.1}  & 100.0 & 102.3 & 102.3 &  $-$0.41 & {\bf $-$0.95}  & $-$1.17 & $-$1.21 & $-$1.21  &  +0.11 &  +0.34 \\
6500 & 2.0 & $-$4.0 &  2.0 &  3.71 &  9212.86 &    3.5 &    9.6 & {\bf  19.5}  &  24.0 &  25.1 &  25.1 &  $-$0.45 & {\bf $-$0.80}  & $-$0.91 & $-$0.93 & $-$0.93  &  +0.15 &  +0.13 \\
6500 & 4.0 &  0.0 &  2.0 &  7.21 &  9212.86 &  186.2 &  204.2 & {\bf 234.4}  & 263.0 & 275.4 & 281.8 &  $-$0.12 & {\bf $-$0.31}  & $-$0.49 & $-$0.56 & $-$0.57  &   0.00 &  +0.02 \\
6500 & 4.0 & $-$1.0 &  2.0 &  6.71 &  9212.86 &  134.9 &  147.9 & {\bf 177.8}  & 213.8 & 229.1 & 234.4 &  $-$0.13 & {\bf $-$0.40}  & $-$0.66 & $-$0.78 & $-$0.80  &  $-$0.01 &  +0.03 \\
6500 & 4.0 & $-$2.0 &  2.0 &  5.71 &  9212.86 &   58.9 &   63.1 & {\bf  77.6}  & 104.7 & 125.9 & 128.8 &  $-$0.06 & {\bf $-$0.26}  & $-$0.64 & $-$0.90 & $-$0.94  &   0.00 &  +0.03 \\
6500 & 4.0 & $-$3.0 &  2.0 &  4.71 &  9212.86 &   11.2 &   12.3 & {\bf  20.0}  &  41.7 &  55.0 &  57.5 &  $-$0.05 & {\bf $-$0.29}  & $-$0.73 & $-$0.94 & $-$0.98  &   0.00 &  +0.05 \\
6500 & 4.0 & $-$4.0 &  2.0 &  3.71 &  9212.86 &    1.2 &    1.6 & {\bf   3.7}  &   8.9 &  12.0 &  12.6 &  $-$0.11 & {\bf $-$0.50}  & $-$0.89 & $-$1.03 & $-$1.06  &  +0.04 &  +0.01 \\
\hline
\end{tabular}
\end{center}
Columns 1--6 are self-explanatory (the units of $T_{\rm eff}$, $g$, 
and $\xi$ are K, cm~s$^{-2}$, and km~s$^{-1}$, respectively). 
While $W^{\rm LTE}$ in the 7th column is the LTE equivalent width
(calculated for the atmospheric parameters and the input abundance 
given in columns 1--5), the $W$s in columns 8--12 and $\Delta$s in
columns 13-17 are the non-LTE equivalent width (in m$\rm\AA$) and 
the non-LTE abundance corrections (in dex), respectively, where the 
suffixes ($+1$, $0$, $-1$, $-2$, $-3$, and $-4$) denote the corresponding 
values of $h$ (the logarithm of the H~{\sc i} collision correction 
factor applied to the classical formula).
The values for the finally adopted $h=0$ case are highlighted by 
{\bf boldface} characters. 
In the 18th and 19th columns are given the variation of $\Delta_{0}$ 
caused by chaging the photoionization cross section (for all levels)
by a factor of $1/10$ 
[$\delta(\Delta_{-}) \equiv \Delta_{0}(\alpha_{\rm std}) - \Delta_{0}(\alpha_{\rm std} \times 0.1)$] and 10 [$\delta(\Delta_{+}) \equiv 
\Delta_{0}(\alpha_{\rm std}) - \Delta_{0}(\alpha_{\rm std} \times 10)$].

\end{table}

\clearpage
\setcounter{table}{2}
\begin{table}[h]
\tiny
\caption{Dependence of the non-LTE effect of Zn~{\sc i} 4810 and 6362 lines
on the H~{\sc i} collision and photoionization cross section.} 
\begin{center}
\begin{tabular}
{ccrcccrr@{ }r@{ }r@{ }r@{ }rr@{ }r@{ }r@{ }r@{ }rr@{}r}\hline \hline
$T_{\rm eff}$  & $\log g$ & [Fe/H] & $\xi$ & $A_{\rm Zn}$ & Line &
$W^{\rm LTE}$ & $W_{+1}$ & $W_{0}$ & $W_{-1}$ & $W_{-2}$ & $W_{-3}$ &
$\Delta_{+1}$ & $\Delta_{0}$ & $\Delta_{-1}$ & $\Delta_{-2}$ & $\Delta_{-3}$ 
& $\delta(\Delta)_{-}$ & $\delta(\Delta)_{+}$\\
\hline
4500 & 2.0 &  0.0 &  2.0 &  4.60 &  4810.53 &  114.8 &  120.2 & {\bf 123.0}  & 128.8 & 131.8 & 134.9 &  $-$0.08 & {\bf $-$0.17}  & $-$0.28 & $-$0.35 & $-$0.37  &   0.00 &   0.00 \\
4500 & 2.0 & $-$1.0 &  2.0 &  3.60 &  4810.53 &   77.6 &  $\cdots$ & {\bf $\cdots$}  & $\cdots$ & $\cdots$ & $\cdots$ & $\cdots$ & {\bf $\cdots$}  & $\cdots$ &  $\cdots$ &  $\cdots$  &  $\cdots$ &  $\cdots$ \\
4500 & 2.0 & $-$2.0 &  2.0 &  2.60 &  4810.53 &   33.9 &   34.7 & {\bf  39.8}  &  42.7 &  41.7 &  40.7 &  $-$0.01 & {\bf $-$0.11}  & $-$0.16 & $-$0.14 & $-$0.14  &  +0.02 &  +0.03 \\
4500 & 2.0 & $-$3.0 &  2.0 &  2.10 &  4810.53 &   15.9 &   16.2 & {\bf  18.2}  &  17.8 &  16.6 &  15.9 &  $-$0.02 & {\bf $-$0.07}  & $-$0.07 & $-$0.03 & $-$0.01  &   0.00 &  +0.01 \\
4500 & 2.0 & $-$4.0 &  2.0 &  1.10 &  4810.53 &    1.9 &    2.3 & {\bf   2.3}  &   2.0 &   1.8 &   1.7 &  $-$0.09 & {\bf $-$0.09}  & $-$0.04 &  0.02 &  0.04  &   0.00 &  +0.02 \\
4500 & 4.0 &  0.0 &  2.0 &  4.60 &  4810.53 &   83.2 &   83.2 & {\bf  85.1}  &  89.1 &  95.5 & 100.0 &   0.00 & {\bf $-$0.01}  & $-$0.09 & $-$0.19 & $-$0.25  &   0.00 &   0.00 \\
4500 & 4.0 & $-$1.0 &  2.0 &  3.60 &  4810.53 &   41.7 &   41.7 & {\bf  41.7}  &  47.9 &  55.0 &  58.9 &   0.00 & {\bf $-$0.01}  & $-$0.11 & $-$0.23 & $-$0.28  &  +0.01 &  +0.01 \\
4500 & 4.0 & $-$2.0 &  2.0 &  2.60 &  4810.53 &   10.5 &   10.7 & {\bf  11.8}  &  15.5 &  18.2 &  18.2 &  $-$0.01 & {\bf $-$0.05}  & $-$0.19 & $-$0.28 & $-$0.27  &  +0.04 &  +0.04 \\
4500 & 4.0 & $-$3.0 &  2.0 &  2.10 &  4810.53 &    4.5 &    4.6 & {\bf   5.0}  &   6.5 &   7.2 &   6.9 &  $-$0.01 & {\bf $-$0.05}  & $-$0.17 & $-$0.22 & $-$0.20  &   0.00 &   0.00 \\
4500 & 4.0 & $-$4.0 &  2.0 &  1.10 &  4810.53 &    0.5 &    0.5 & {\bf   0.6}  &   0.8 &   0.8 &   0.7 &  $-$0.03 & {\bf $-$0.12}  & $-$0.23 & $-$0.22 & $-$0.18  &   0.00 &   0.00 \\
5500 & 2.0 &  0.0 &  2.0 &  4.60 &  4810.53 &  117.5 &  123.0 & {\bf 123.0}  & 123.0 & 125.9 & 125.9 &  $-$0.10 & {\bf $-$0.09}  & $-$0.11 & $-$0.13 & $-$0.14  &  +0.01 &   0.00 \\
5500 & 2.0 & $-$1.0 &  2.0 &  3.60 &  4810.53 &   67.6 &   66.1 & {\bf  63.1}  &  58.9 &  57.5 &  57.5 &   0.05 & {\bf  0.10}  &  0.17 &  0.20 &  0.21  &  $-$0.16 &  $-$0.05 \\
5500 & 2.0 & $-$2.0 &  2.0 &  2.60 &  4810.53 &   18.2 &   15.1 & {\bf  13.5}  &  11.0 &  10.0 &   9.8 &   0.09 & {\bf  0.16}  &  0.26 &  0.31 &  0.31  &  $-$0.03 &  +0.09 \\
5500 & 2.0 & $-$3.0 &  2.0 &  2.10 &  4810.53 &    6.9 &    5.5 & {\bf   4.5}  &   3.5 &   3.2 &   3.1 &   0.10 & {\bf  0.20}  &  0.32 &  0.36 &  0.37  &  $-$0.03 &  +0.09 \\
5500 & 2.0 & $-$4.0 &  2.0 &  1.10 &  4810.53 &    0.8 &    0.6 & {\bf   0.4}  &   0.3 &   0.3 &   0.3 &   0.12 & {\bf  0.25}  &  0.36 &  0.39 &  0.40  &  $-$0.02 &  +0.07 \\
5500 & 4.0 &  0.0 &  2.0 &  4.60 &  4810.53 &  102.3 &  104.7 & {\bf 107.2}  & 109.7 & 112.2 & 114.8 &  $-$0.03 & {\bf $-$0.07}  & $-$0.12 & $-$0.18 & $-$0.20  &   0.00 &  +0.01 \\
5500 & 4.0 & $-$1.0 &  2.0 &  3.60 &  4810.53 &   53.7 &   52.5 & {\bf  52.5}  &  55.0 &  55.0 &  53.7 &   0.01 & {\bf  0.01}  & $-$0.03 & $-$0.02 &  0.00  &  $-$0.05 &  $-$0.02 \\
5500 & 4.0 & $-$2.0 &  2.0 &  2.60 &  4810.53 &   10.7 &   10.5 & {\bf  10.7}  &  11.0 &   9.3 &   8.3 &   0.01 & {\bf  0.00}  & $-$0.01 &  0.07 &  0.12  &  +0.01 &  +0.05 \\
5500 & 4.0 & $-$3.0 &  2.0 &  2.10 &  4810.53 &    3.8 &    3.9 & {\bf   4.1}  &   3.8 &   3.0 &   2.6 &   0.00 & {\bf $-$0.03}  &  0.01 &  0.12 &  0.18  &   0.00 &  +0.03 \\
5500 & 4.0 & $-$4.0 &  2.0 &  1.10 &  4810.53 &    0.4 &    0.5 & {\bf   0.5}  &   0.3 &   0.3 &   0.2 &  $-$0.07 & {\bf $-$0.06}  &  0.05 &  0.17 &  0.22  &   0.00 &  +0.04 \\
6500 & 2.0 &  0.0 &  2.0 &  4.60 &  4810.53 &   91.2 &   91.2 & {\bf  89.1}  &  87.1 &  87.1 &  87.1 &  $-$0.02 & {\bf  0.04}  &  0.07 &  0.08 &  0.08  &  $-$0.02 &  +0.04 \\
6500 & 2.0 & $-$1.0 &  2.0 &  3.60 &  4810.53 &   36.3 &   30.2 & {\bf  26.9}  &  24.5 &  24.0 &  24.0 &   0.13 & {\bf  0.20}  &  0.25 &  0.26 &  0.26  &  $-$0.09 &  +0.03 \\
6500 & 2.0 & $-$2.0 &  2.0 &  2.60 &  4810.53 &    5.8 &    3.7 & {\bf   3.0}  &   2.7 &   2.7 &   2.7 &   0.21 & {\bf  0.30}  &  0.34 &  0.35 &  0.35  &  $-$0.03 &  +0.01 \\
6500 & 2.0 & $-$3.0 &  2.0 &  2.10 &  4810.53 &    1.5 &    0.9 & {\bf   0.8}  &   0.7 &   0.7 &   0.7 &   0.21 & {\bf  0.28}  &  0.31 &  0.31 &  0.30  &  +0.03 &  $-$0.02 \\
6500 & 2.0 & $-$4.0 &  2.0 &  1.10 &  4810.53 &    0.2 &    0.1 & {\bf   0.1}  &   0.1 &   0.1 &   0.1 &   0.22 & {\bf  0.29}  &  0.32 &  0.32 &  0.32  &  +0.03 &  $-$0.02 \\
6500 & 4.0 &  0.0 &  2.0 &  4.60 &  4810.53 &   83.2 &   85.1 & {\bf  85.1}  &  83.2 &  83.2 &  83.2 &  $-$0.04 & {\bf $-$0.03}  & $-$0.01 &  0.00 &  0.00  &  $-$0.01 &  +0.06 \\
6500 & 4.0 & $-$1.0 &  2.0 &  3.60 &  4810.53 &   30.2 &   28.2 & {\bf  26.3}  &  24.5 &  22.9 &  22.4 &   0.04 & {\bf  0.08}  &  0.12 &  0.17 &  0.18  &  $-$0.06 &  +0.05 \\
6500 & 4.0 & $-$2.0 &  2.0 &  2.60 &  4810.53 &    4.6 &    4.1 & {\bf   3.6}  &   3.0 &   2.7 &   2.6 &   0.05 & {\bf  0.10}  &  0.18 &  0.23 &  0.24  &  $-$0.02 &  +0.06 \\
6500 & 4.0 & $-$3.0 &  2.0 &  2.10 &  4810.53 &    1.5 &    1.4 & {\bf   1.1}  &   0.9 &   0.8 &   0.8 &   0.05 & {\bf  0.14}  &  0.23 &  0.29 &  0.30  &  $-$0.01 &  +0.06 \\
6500 & 4.0 & $-$4.0 &  2.0 &  1.10 &  4810.53 &    0.2 &    0.1 & {\bf   0.1}  &   0.1 &   0.1 &   0.1 &   0.04 & {\bf  0.16}  &  0.24 &  0.30 &  0.32  &  $-$0.01 &  +0.05 \\
\hline
4500 & 2.0 &  0.0 &  2.0 &  4.60 &  6362.34 &   28.8 &   28.8 & {\bf  30.9}  &  33.9 &  33.1 &  33.1 &   0.00 & {\bf $-$0.05}  & $-$0.12 & $-$0.11 & $-$0.10  &   0.00 &   0.00 \\
4500 & 2.0 & $-$1.0 &  2.0 &  3.60 &  6362.34 &    7.4 & $\cdots$ & {\bf $\cdots$}  & $\cdots$ & $\cdots$ & $\cdots$ & $\cdots$ & {\bf $\cdots$}  & $\cdots$ &  $\cdots$ & $\cdots$  &  $\cdots$ &  $\cdots$ \\
4500 & 2.0 & $-$2.0 &  2.0 &  2.60 &  6362.34 &    1.1 &    1.0 & {\bf   1.2}  &   2.0 &   2.4 &   2.5 &   0.03 & {\bf $-$0.05}  & $-$0.27 & $-$0.35 & $-$0.37  &  +0.04 &  +0.04 \\
4500 & 2.0 & $-$3.0 &  2.0 &  2.10 &  6362.34 &    0.4 &    0.3 & {\bf   0.4}  &   0.6 &   0.7 &   0.7 &   0.04 & {\bf $-$0.03}  & $-$0.26 & $-$0.30 & $-$0.30  &   0.00 &  +0.01 \\
4500 & 2.0 & $-$4.0 &  2.0 &  1.10 &  6362.34 &    0.0 &    0.0 & {\bf   0.1}  &   0.1 &   0.1 &   0.1 &  $-$0.02 & {\bf $-$0.13}  & $-$0.29 & $-$0.26 & $-$0.24  &   0.00 &  +0.02 \\
4500 & 4.0 &  0.0 &  2.0 &  4.60 &  6362.34 &    8.9 &    8.9 & {\bf   8.9}  &   9.3 &  10.0 &   9.8 &   0.00 & {\bf  0.00}  & $-$0.03 & $-$0.06 & $-$0.04  &   0.00 &   0.00 \\
4500 & 4.0 & $-$1.0 &  2.0 &  3.60 &  6362.34 &    1.7 &    1.7 & {\bf   1.7}  &   1.9 &   2.5 &   2.6 &   0.00 & {\bf  0.01}  & $-$0.04 & $-$0.16 & $-$0.19  &  $-$0.01 &   0.00 \\
4500 & 4.0 & $-$2.0 &  2.0 &  2.60 &  6362.34 &    0.2 &    0.2 & {\bf   0.2}  &   0.3 &   0.6 &   0.8 &   0.00 & {\bf  0.00}  & $-$0.12 & $-$0.38 & $-$0.51  &  +0.01 &  +0.01 \\
4500 & 4.0 & $-$3.0 &  2.0 &  2.10 &  6362.34 &    0.1 &    0.1 & {\bf   0.1}  &   0.1 &   0.2 &   0.3 &   0.00 & {\bf $-$0.02}  & $-$0.14 & $-$0.37 & $-$0.46  &   0.00 &   0.00 \\
4500 & 4.0 & $-$4.0 &  2.0 &  1.10 &  6362.34 &    0.0 &    0.0 & {\bf   0.0}  &   0.0 &   0.0 &   0.0 &  $-$0.01 & {\bf $-$0.09}  & $-$0.26 & $-$0.42 & $-$0.43  &   0.00 &   0.00 \\
5500 & 2.0 &  0.0 &  2.0 &  4.60 &  6362.34 &   46.8 &   46.8 & {\bf  49.0}  &  51.3 &  50.1 &  50.1 &  $-$0.01 & {\bf $-$0.05}  & $-$0.08 & $-$0.06 & $-$0.06  &   0.00 &   0.00 \\
5500 & 2.0 & $-$1.0 &  2.0 &  3.60 &  6362.34 &    8.3 &    7.8 & {\bf   9.3}  &  11.2 &  12.0 &  12.0 &   0.03 & {\bf $-$0.05}  & $-$0.14 & $-$0.17 & $-$0.18  &   0.00 &  +0.06 \\
5500 & 2.0 & $-$2.0 &  2.0 &  2.60 &  6362.34 &    0.9 &    0.9 & {\bf   1.2}  &   1.3 &   1.3 &   1.3 &   0.01 & {\bf $-$0.12}  & $-$0.15 & $-$0.14 & $-$0.14  &  +0.01 &  +0.11 \\
5500 & 2.0 & $-$3.0 &  2.0 &  2.10 &  6362.34 &    0.3 &    0.3 & {\bf   0.3}  &   0.4 &   0.4 &   0.4 &   0.02 & {\bf $-$0.05}  & $-$0.09 & $-$0.09 & $-$0.08  &  $-$0.04 &  +0.09 \\
5500 & 2.0 & $-$4.0 &  2.0 &  1.10 &  6362.34 &    0.0 &    0.0 & {\bf   0.0}  &   0.0 &   0.0 &   0.0 &   0.06 & {\bf  0.05}  & $-$0.02 & $-$0.04 & $-$0.05  &  $-$0.03 &  +0.08 \\
5500 & 4.0 &  0.0 &  2.0 &  4.60 &  6362.34 &   26.9 &   27.5 & {\bf  27.5}  &  29.5 &  30.2 &  29.5 &   0.00 & {\bf $-$0.01}  & $-$0.05 & $-$0.07 & $-$0.05  &   0.00 &   0.00 \\
5500 & 4.0 & $-$1.0 &  2.0 &  3.60 &  6362.34 &    4.7 &    4.6 & {\bf   4.5}  &   5.4 &   6.3 &   6.3 &   0.01 & {\bf  0.02}  & $-$0.07 & $-$0.14 & $-$0.14  &  $-$0.02 &  +0.01 \\
5500 & 4.0 & $-$2.0 &  2.0 &  2.60 &  6362.34 &    0.5 &    0.5 & {\bf   0.5}  &   0.7 &   0.9 &   0.8 &   0.02 & {\bf  0.00}  & $-$0.15 & $-$0.22 & $-$0.21  &  +0.02 &  +0.05 \\
5500 & 4.0 & $-$3.0 &  2.0 &  2.10 &  6362.34 &    0.2 &    0.2 & {\bf   0.2}  &   0.2 &   0.2 &   0.2 &   0.01 & {\bf $-$0.02}  & $-$0.13 & $-$0.17 & $-$0.18  &   0.00 &  +0.03 \\
5500 & 4.0 & $-$4.0 &  2.0 &  1.10 &  6362.34 &    0.0 &    0.0 & {\bf   0.0}  &   0.0 &   0.0 &   0.0 &  $-$0.06 & {\bf $-$0.04}  &  0.05 & $-$0.04 & $-$0.10  &   0.00 &  +0.04 \\
6500 & 2.0 &  0.0 &  2.0 &  4.60 &  6362.34 &   26.9 &   28.2 & {\bf  30.2}  &  32.4 &  32.4 &  32.4 &  $-$0.03 & {\bf $-$0.08}  & $-$0.11 & $-$0.12 & $-$0.12  &  $-$0.01 &  +0.01 \\
6500 & 2.0 & $-$1.0 &  2.0 &  3.60 &  6362.34 &    3.9 &    4.1 & {\bf   4.7}  &   4.9 &   5.0 &   5.0 &  $-$0.02 & {\bf $-$0.08}  & $-$0.11 & $-$0.11 & $-$0.11  &  $-$0.04 &  +0.03 \\
6500 & 2.0 & $-$2.0 &  2.0 &  2.60 &  6362.34 &    0.4 &    0.2 & {\bf   0.3}  &   0.4 &   0.4 &   0.4 &   0.25 & {\bf  0.18}  &  0.04 &  0.01 &  0.00  &  $-$0.09 &  $-$0.09 \\
6500 & 2.0 & $-$3.0 &  2.0 &  2.10 &  6362.34 &    0.1 &    0.1 & {\bf   0.1}  &   0.1 &   0.1 &   0.1 &   0.39 & {\bf  0.24}  &  0.10 &  0.06 &  0.05  &  +0.03 &  $-$0.03 \\
6500 & 2.0 & $-$4.0 &  2.0 &  1.10 &  6362.34 &    0.0 &    0.0 & {\bf   0.0}  &   0.0 &   0.0 &   0.0 &   0.45 & {\bf  0.24}  &  0.09 &  0.05 &  0.04  &  +0.03 &  $-$0.03 \\
6500 & 4.0 &  0.0 &  2.0 &  4.60 &  6362.34 &   22.9 &   22.9 & {\bf  23.4}  &  24.0 &  23.4 &  22.9 &   0.00 & {\bf  0.00}  & $-$0.02 &  0.00 &  0.00  &  $-$0.01 &  +0.03 \\
6500 & 4.0 & $-$1.0 &  2.0 &  3.60 &  6362.34 &    3.2 &    3.0 & {\bf   3.1}  &   3.5 &   3.5 &   3.4 &   0.03 & {\bf  0.00}  & $-$0.05 & $-$0.04 & $-$0.04  &  $-$0.01 &  +0.06 \\
6500 & 4.0 & $-$2.0 &  2.0 &  2.60 &  6362.34 &    0.3 &    0.3 & {\bf   0.3}  &   0.3 &   0.3 &   0.3 &   0.05 & {\bf  0.08}  &  0.10 &  0.06 &  0.05  &  $-$0.06 &  +0.02 \\
6500 & 4.0 & $-$3.0 &  2.0 &  2.10 &  6362.34 &    0.1 &    0.1 & {\bf   0.1}  &   0.1 &   0.1 &   0.1 &   0.08 & {\bf  0.29}  &  0.32 &  0.21 &  0.17  &  $-$0.01 &  +0.05 \\
6500 & 4.0 & $-$4.0 &  2.0 &  1.10 &  6362.34 &    0.0 &    0.0 & {\bf   0.0}  &   0.0 &   0.0 &   0.0 &   0.07 & {\bf  0.41}  &  0.34 &  0.18 &  0.13  &  $-$0.01 &  +0.04 \\
\hline
\end{tabular}
\end{center}
See the note in table 2 for a detailed description of the presented data.
The non-LTE $W$s and $\Delta$s could not be successfully calculated 
for the case of ($T_{\rm eff}$ = 4500 K, $\log g$ = 2.0, and [Fe/H] = $-1.0$)
because of instability problems.
\end{table}

\clearpage
\setcounter{table}{3}
\begin{table}[h]
\scriptsize
\caption{Non-LTE analysis of S~{\sc i} and Zn~{\sc i} lines for $\alpha$ Per, Polaris, Procyon, and the Sun.} 
\begin{center}
\begin{tabular}
{crcccrr}\hline \hline
Line & $W_{\lambda}^{*}$ & $\overline{\log \tau}^{\dagger}$ & $A^{\rm NLTE}$ & $A^{\rm LTE}$ & $\Delta^{\ddagger}$ & $\delta_{\rm vdw+}^{\S}$ \\ \hline
\multicolumn{6}{c}{$\alpha$ Per \ ($T_{\rm eff} = 6250$, $\log g = 0.90$, $[X]=0.0$, $\xi = 4.5$) }\\
S~{\sc i} 8693+4 & 264.7 & $-$0.77  & 6.96  & 7.21  & $-$0.25  & 0.00\\
S~{\sc i} 9212 & 521.4 & $-$2.36  & 7.14  & 8.24  & $-$1.10 & $-$0.01\\
S~{\sc i} 9237 & 455.7 & $-$2.18  & 7.22  & 8.31  & $-$1.09 & 0.00\\
\\
Zn~{\sc i} 4722 & 80.1 & $-$0.62  & 4.10  & 4.01  & +0.09 & 0.00\\
Zn~{\sc i} 4810 & 87.8 & $-$0.68  & 4.01  & 3.91  & +0.10 & 0.00\\
Zn~{\sc i} 6362 & 35.8$^{\parallel}$ & $-$0.42$^{\parallel}$  & 4.48$^{\parallel}$  & 4.60$^{\parallel}$  & $-$0.12$^{\parallel}$ & 0.00\\
\hline
\multicolumn{6}{c}{Polaris \ ($T_{\rm eff} = 6000$, $\log g = 1.50$, $[X]=0.0$, $\xi = 5.0$) }\\
S~{\sc i} 8693 & 68.0 & $-$0.50  & 7.04  & 7.19  & $-$0.15 & 0.00\\
S~{\sc i} 8694 & 128.2 & $-$0.70  & 6.87  & 7.10  & $-$0.23 & 0.00\\
S~{\sc i} 9212 & 458.6 & $-$2.04  & 6.87  & 7.80  & $-$0.93 & 0.00\\
S~{\sc i} 9237 & 352.9 & $-$1.70  & 6.82  & 7.50  & $-$0.68 & 0.00\\
\\
Zn~{\sc i} 4722 & 99.8 & $-$0.76  & 4.13  & 4.07  & +0.06 & 0.00\\
Zn~{\sc i} 4810 & 110.5 & $-$0.83  & 4.05  & 3.99  & +0.06 & 0.00\\
Zn~{\sc i} 6362 & 28.8 & $-$0.37  & 4.34  & 4.41  & $-$0.07 & 0.00\\
\hline
\multicolumn{6}{c}{Procyon \ ($T_{\rm eff} = 6600$, $\log g = 4.00$, $[X]=0.0$, $\xi = 2.0$) }\\
S~{\sc i} 8693 & 28.8 & $-$0.51  & 7.18  & 7.22  & $-$0.04 & $-$0.01\\
S~{\sc i} 8694 & 72.2 & $-$0.78  & 7.24  & 7.31  & $-$0.07 & $-$0.04\\
S~{\sc i} 9212 & 207.7 & $-$1.93  & 7.08  & 7.47  & $-$0.39 & $-$0.10\\
S~{\sc i} 9237 & 161.8 & $-$1.74  & 7.16  & 7.54  & $-$0.38 & $-$0.07\\
\\
Zn~{\sc i} 4722 & 65.2 & $-$0.94  & 4.46  & 4.49  & $-$0.03 & $-$0.01\\
Zn~{\sc i} 4810 & 71.0 & $-$1.03  & 4.38  & 4.44  & $-$0.06 & $-$0.01\\
Zn~{\sc i} 6362 & 18.8 & $-$0.40  & 4.46  & 4.52  & $-$0.06 & 0.00\\
\hline
\multicolumn{6}{c}{Sun \ ($T_{\rm eff} = 5780$, $\log g = 4.44$, $[X]=0.0$, $\xi = 1.0$) }\\
S~{\sc i} 8693 & 10.6 & $-$0.36  & 7.18  & 7.19  & $-$0.01 & $-$0.03\\
S~{\sc i} 8694 & 28.5 & $-$0.45  & 7.17  & 7.18  & $-$0.01 & $-$0.06\\
S~{\sc i} 9228 & 95.1 & $-$1.07  & 7.06  & 7.21  & $-$0.15 & $-$0.15\\
S~{\sc i} 9237 & 97.1 & $-$0.95  & 7.14  & 7.25  & $-$0.11 & $-$0.16\\
S~{\sc i} 10455 & 111.9 & $-$0.82  & 7.11  & 7.20  & $-$0.09 & $-$0.16\\
S~{\sc i} 10456 & 55.3 & $-$0.58  & 7.14  & 7.19  & $-$0.05 & $-$0.08\\
S~{\sc i} 10459 & 88.2 & $-$0.73  & 7.09  & 7.16  & $-$0.07 & $-$0.10\\
\\
Zn~{\sc i} 4722 & 67.4 & $-$1.23  & 4.56  & 4.61  & $-$0.05 & $-$0.13\\
Zn~{\sc i} 4810 & 71.6 & $-$1.28  & 4.49  & 4.54  & $-$0.05 & $-$0.16\\
Zn~{\sc i} 6362 & 20.5 & $-$0.50  & 4.53  & 4.53  & 0.00 & $-$0.03\\
\hline
\end{tabular}
\end{center}
$^{*}$ Equivalent width in units of m$\rm\AA$.\\
$^{\dagger}$ Mean line-formation depth in terms of the standard continuum
optical depth at 5000~$\rm\AA$, which was calculated (for the non-LTE case)
in the same manner as described in Takeda and Takada-Hidai (1994).\\
$^{\ddagger}$ Non-LTE correction defined as 
$\Delta \equiv A^{\rm NLTE} - A^{\rm LTE}$.\\
$^{\S}$ Variation of $A^{\rm NLTE}$ caused by increasing the van der Waals
damping width (for which we adopted the default treatment of the WIDTH9 
program, equivalent to the classical Uns\"{o}ld's approximation; cf. table 1) 
by a factor of 2.5.\\
$^{\parallel}$ These values are less reliable and should be viewed with caution
because of the difficulty in measuring the equivalent width (cf. figure 3c).
\end{table}

\clearpage

\onecolumn

\begin{figure}
  \begin{center}
    \FigureFile(140mm,140mm){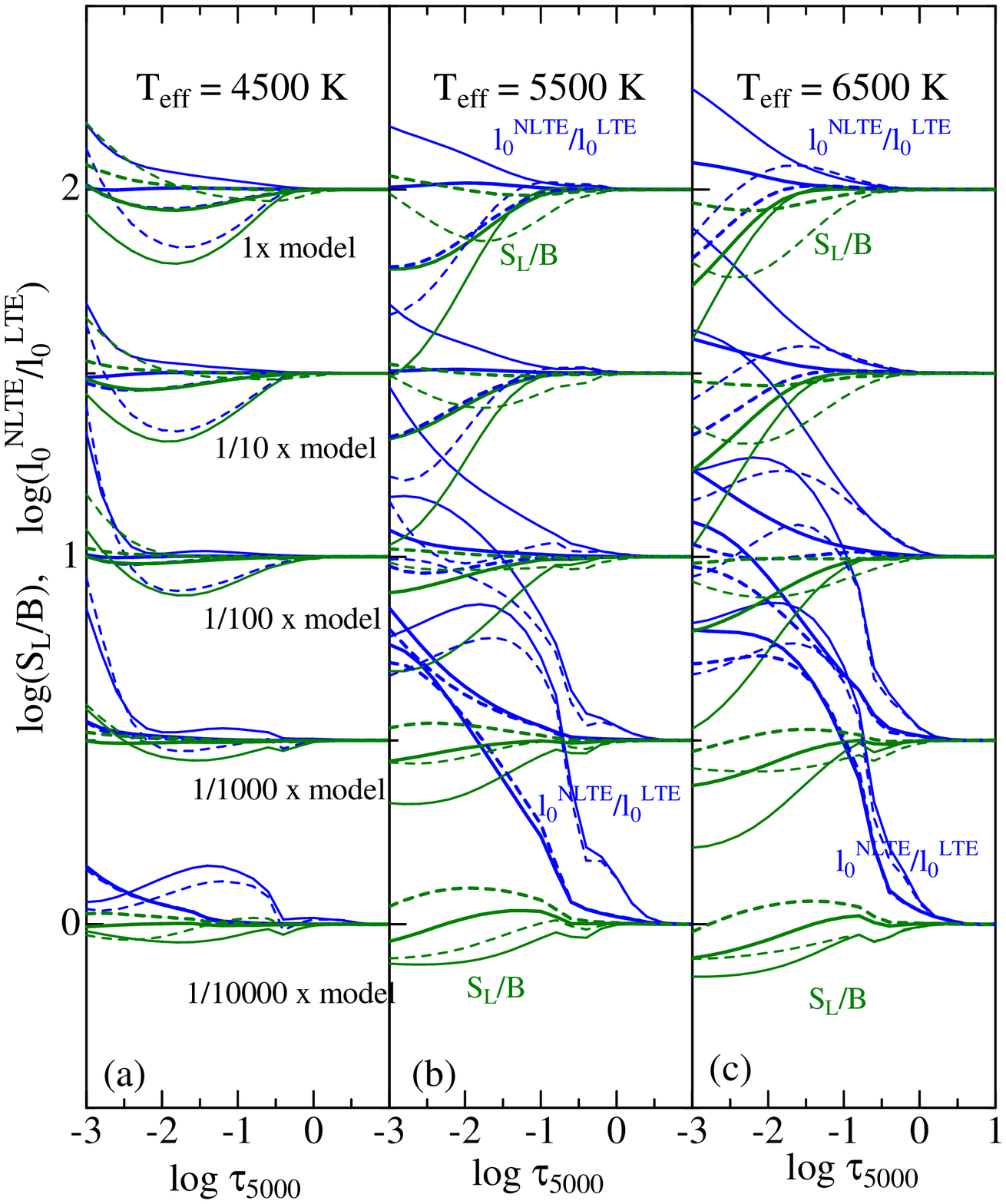}
  \end{center}
\caption{
Ratio of the S~{\sc i} line source function ($S_{\rm L}$) to the local
Planck function ($B$) and the NLTE-to-LTE line-center opacity
ratio as functions of the standard continuum optical depth
at 5000 $\rm\AA$ computed for models of $T_{\rm eff}$ = 4500~K, 5500~K, 
and 6500~K. The green lines and blue lines correspond to $S_{\rm L}/B$ 
and $l_{0}^{\rm NLTE}/l_{0}^{\rm LTE}$, respectively.
The solid lines show the results for the 
4s $^{5}{\rm S}^{\rm o}$ -- 4p $^{5}{\rm P}$
transition of  multiplet 1 (corresponding to S~{\sc i} 9212/9228/9237 
lines), while those for the 
4p $^{5}{\rm P}$ -- 4d $^{5}{\rm D}^{\rm o}$ transition
of multiplet 6 (corresponding to S~{\sc i} 8693/8694 lines) are depicted
by dashed lines. In each case, the results for two different gravity
atmospheres are given: The thick lines are for $\log g = 4$ and the thin lines
are for $\log g = 2$. Note also that the curves are 
vertically offset by an amount of 0.5~dex relative to those of the 
adjacent metallicity ones. 
}
\end{figure}

\begin{figure}
  \begin{center}
    \FigureFile(140mm,140mm){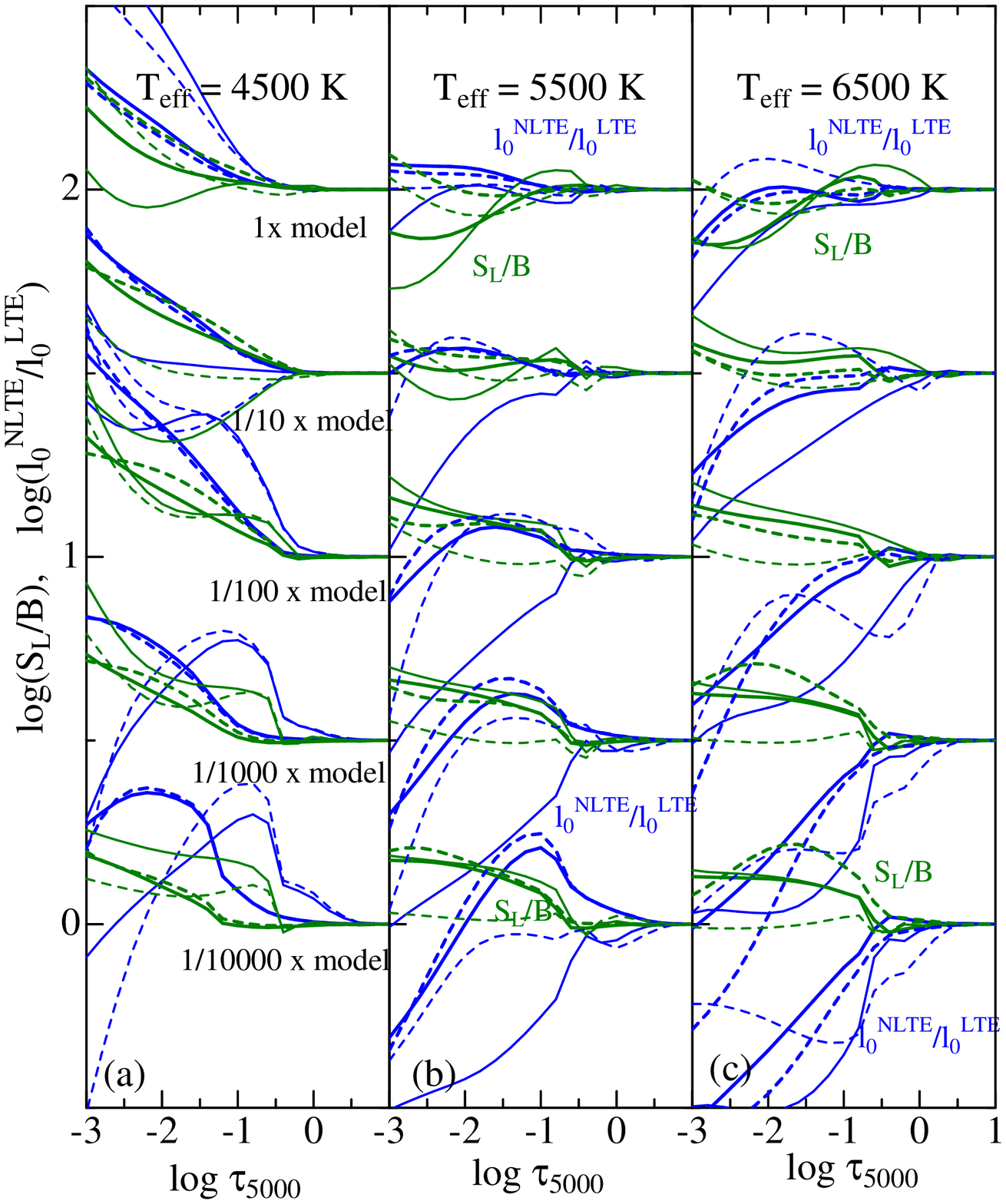}
  \end{center}
\caption{
Ratio of the Zn~{\sc i} line source function ($S_{\rm L}$) to the local
Planck function ($B$) and the NLTE-to-LTE line-center opacity
ratio as functions of the standard continuum optical depth
at 5000 $\rm\AA$ computed for models of $T_{\rm eff}$ = 4500~K, 5500~K, 
and 6500~K. The green lines and blue lines correspond to $S_{\rm L}/B$ 
and $l_{0}^{\rm NLTE}/l_{0}^{\rm LTE}$, respectively.
The solid lines show the results for the 
4p $^{3}{\rm P}^{\rm o}$ -- 5s $^{3}{\rm S}$ 
transition of  multiplet 2 (corresponding to Zn~{\sc i} 4722/4810 
lines), while those for the transition
4p $^{1}{\rm P}^{\rm o}$ -- 4d $^{1}{\rm D}$ of multiplet 6 
(corresponding to Zn~{\sc i} 6362 line) are depicted
by dashed lines. In each case, the results for two different gravity
atmospheres are given: The thick lines are for $\log g = 4$ and the thin lines
are for $\log g = 2$. Note also that the curves are 
vertically offset by an amount of 0.5~dex relative to those of the 
adjacent metallicity ones. 
}
\end{figure}

\begin{figure}
  \begin{center}
    \FigureFile(155mm,155mm){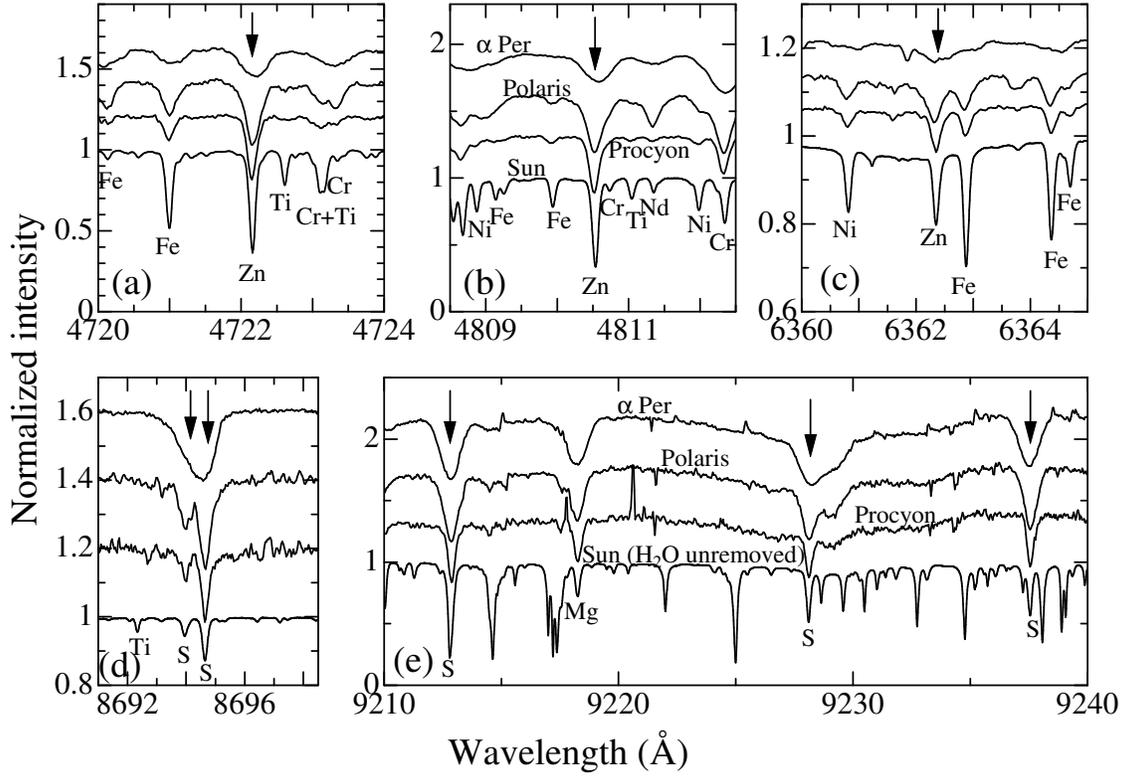}
  \end{center}
\caption{Spectra of three bright F stars observed by using the GAOES 
spectrograph at Gunma Astrophysical Observatory at five wavelength 
regions (a---Zn~{\sc i} 4722 line region, b---Zn~{\sc i} 4810 line 
region, c---Zn~{\sc i} 6362 line region, d---S~{\sc i} 8693/8694 lines
region, e---S~{\sc i} 9212/9228/9237 lines region), on which 
the equivalent widths of these S~{\sc i} and Zn~{\sc i} lines 
(their positions are indicated by downward arrows) were
measured for an adequacy check of our non-LTE calculations. 
The KPNO solar flux spectra of Kurucz et al. (1984) are also shown 
for comparison. The spectra are placed according to the order 
of $\alpha$ Per, Polaris, Procyon, and the Sun from top to bottom,
each being vertically offset by an appropriate constant 
(0.2, 0.3, 0.075, 0.2, and 0.4 for panels a, b, c, d, and e,
respectively) relative to those of the adjacent metallicity ones. 
Note that, in panel (e) of 9210--9240 $\rm\AA$ region, 
numerous telluric lines due to H$_{2}$O have been removed for 
the GAOES spectra by dividing them by the spectrum of a rapid rotator 
($\gamma$ Cas), unlike the KPNO solar spectrum where those telluric
lines are conspicuously observed.}
\end{figure}

\begin{figure}
  \begin{center}
    \FigureFile(140mm,140mm){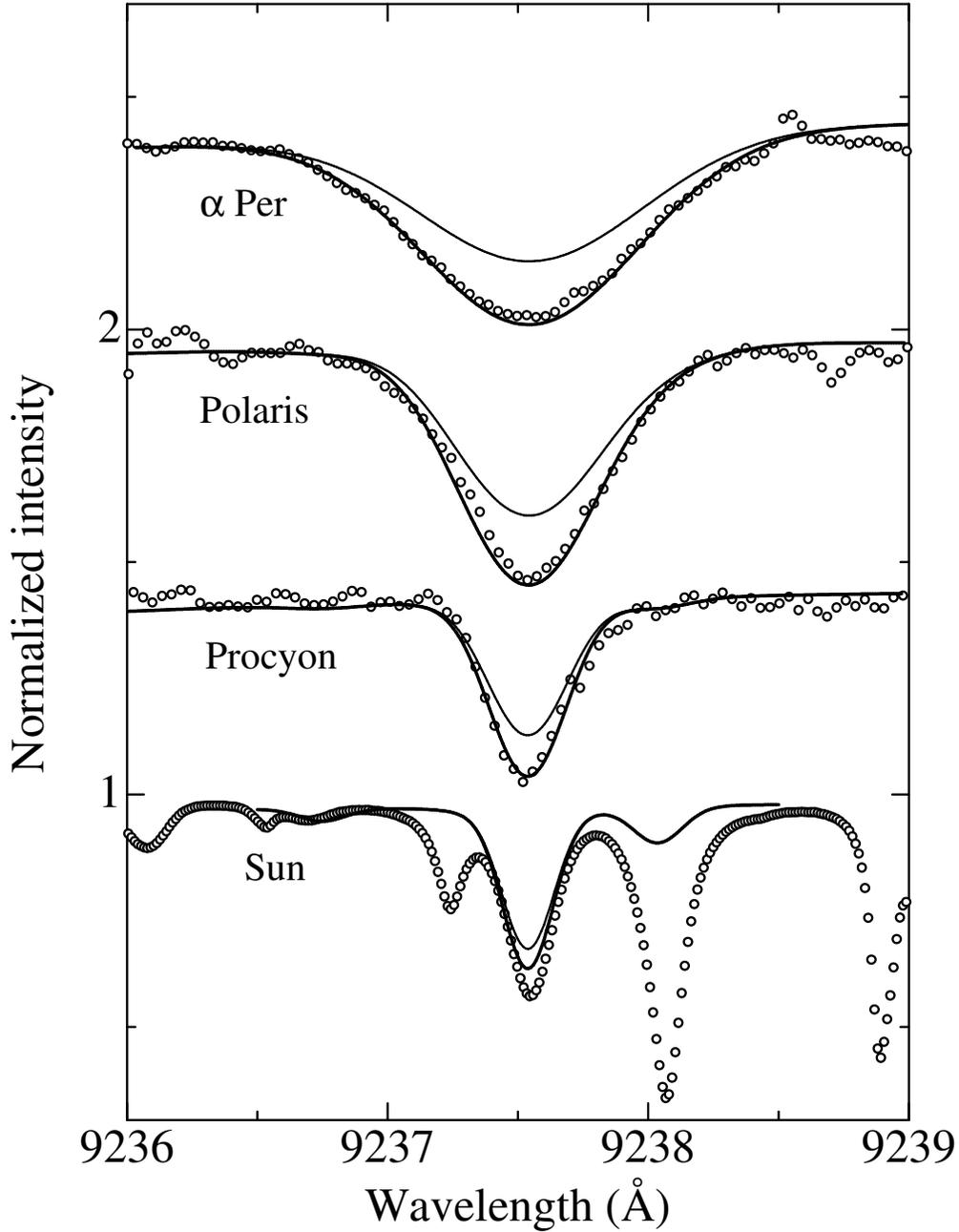}
  \end{center}
\caption{Observed profiles (open circles; the data are the same as figure 3) 
of the S~{\sc i} line at 9237.538 $\rm\AA$ for $\alpha$ Per, 
Polaris, Procyon, and the Sun (from top to bottom; each spectrum is 
vertically shifted by 0.5 relative to the adjacent one), fitted with 
the theoretically calculated profiles (solid lines). 
The theoretical profiles were computed with the non-LTE abundances 
($A^{\rm NLTE}$) derived from the equivalent-width analysis along with 
the atmospheric parameters presented in table 4, and then convolved 
with Gaussian broadening functions appropriately chosen so as to 
make the best fit. (The continuum levels and the wavelength scales 
of the observed spectra have also been so adequately adjusted as to 
accomplish the best match.) In addition to the non-LTE profiles depicted 
in thick lines, the corresponding LTE profiles (computed also with 
$A^{\rm NLTE}$) are also shown in thin lines. For the case of the Sun, 
since the overlapping wings of telluric lines are not included in 
our spectral synthesis, the fit does not appear to be satisfactorily 
good. The weak absorption feature at $\lambda \sim 9238 \rm\AA$ 
recognized in the theoretical solar spectrum is due to the 
Si~{\sc i} 9238.04 line, which is also blended with a strong telluric line.}
\end{figure}

\begin{figure}
  \begin{center}
    \FigureFile(120mm,120mm){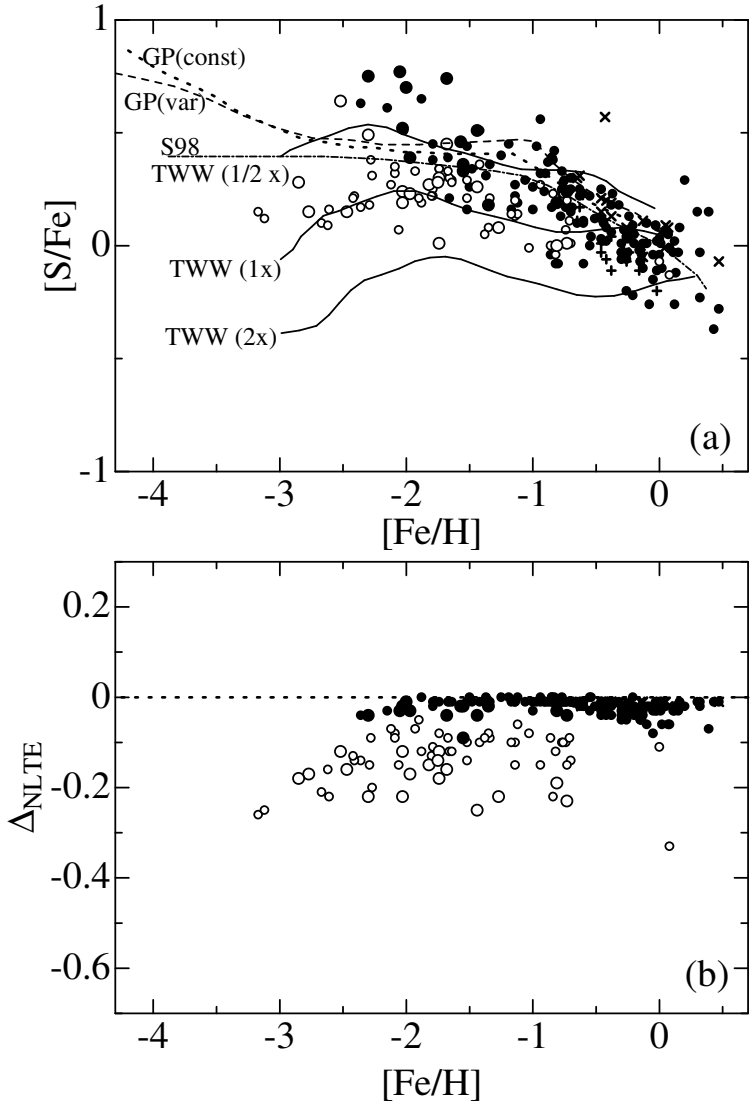}
  \end{center}
\caption{(a) [S/Fe] vs. [Fe/H] relation 
resulting from our non-LTE reanalysis of the published equivalent-width
data of S~{\sc i} lines taken from various literature. 
Open circles --- results from S~{\sc i} 9212/9237 lines of multiplet 1;
filled circles --- results from S~{\sc i} 8693/8694 lines of multiplet 6;
Greek crosses (+) --- results from S~{\sc i} 6756 line of multiplet 8; 
St. Andrew's crosses (x) --- results from S~{\sc i} 6052 line of multiplet 10.
Note that the larger symbol corresponds to low-gravity giant stars 
($\log g < 3$) and the smaller symbol corresponds to high-gravity 
dwarf/subgiant stars ($\log g > 3$).
The representative theoretical predictions are depicted by lines: 
Dash-dotted line (S98)--- taken from figure 12 of Samland (1998);  
solid lines (TWW 1/2~x, 1x, 2x) --- taken from figure 22 of 
Timmes, Woosley, and Weaver (1995) corresponding to three choices of the
adjustment factor (1/2, 1, and 2) for the Fe yield from massive stars 
by which the standard Woosley and Weaver's (1995) yield is to be multiplied; 
dashed/dotted lines --- taken from figure 7 of Goswami and Prantzos (2000) 
for the two cases of S yield, i.e., the dotted line is for the 
metallicity-independent yield [GP(const)] and the dashed line is for the 
realistic metallicity-dependent yield [GP(var)]. 
(b) The corresponding non-LTE corrections used for deriving the 
[S/Fe] values shown in panel (a), plotted as functions of [Fe/H].
}
\end{figure}

\begin{figure}
  \begin{center}
    \FigureFile(120mm,120mm){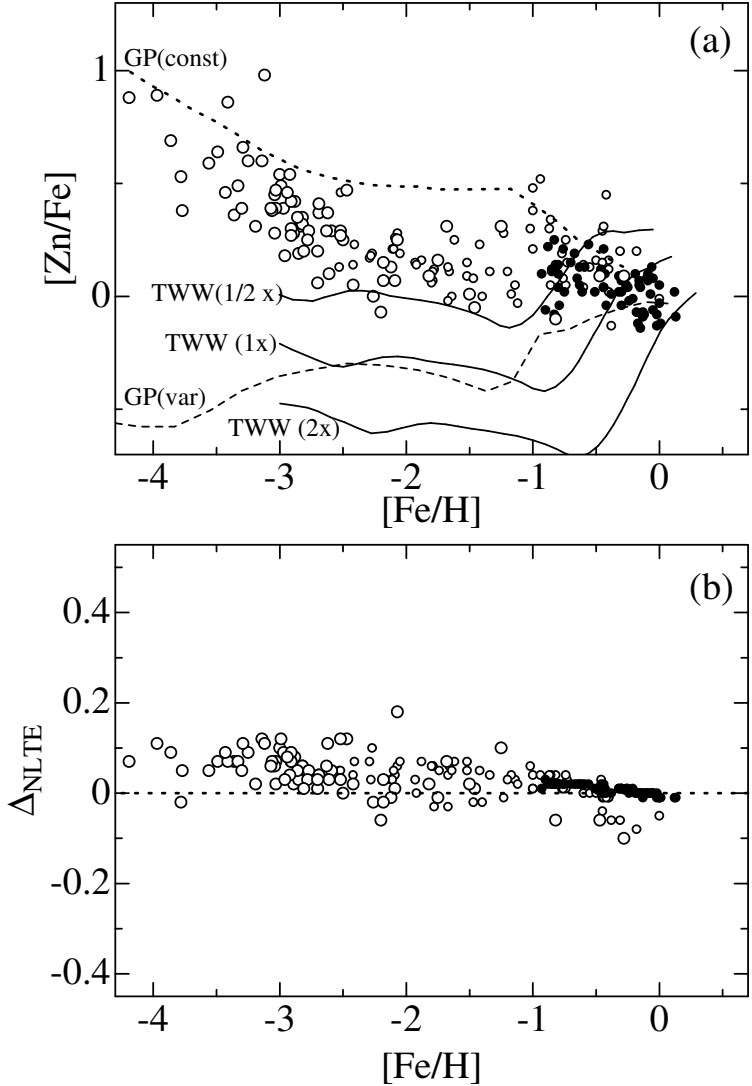}
  \end{center}
\caption{
(a) [Zn/Fe] vs. [Fe/H] relation 
resulting from our non-LTE reanalysis of the published equivalent-width
data of Zn~{\sc i} lines taken from various literature. 
Open circles --- results from Zn~{\sc i} 4722/4780 lines of multiplet 2;
filled circles --- results from Zn~{\sc i} 6362 line of multiplet 6;
Note that the larger symbol corresponds to low-gravity giant stars 
($\log g < 3$) and the smaller symbol corresponds to high-gravity 
dwarf/subgiant stars ($\log g > 3$).
The representative theoretical predictions are depicted by lines: 
Solid lines (TWW 1/2~x, 1x, 2x) --- taken from figure 35 of 
Timmes, Woosley, and Weaver (1995) corresponding to three choices of the
adjustment factor (1/2, 1, and 2) for the Fe yield from massive stars 
by which the standard Woosley and Weaver's (1995) yield is to be multiplied; 
dashed/dotted lines --- taken from figure 7 of Goswami and Prantzos (2000) 
for the two cases of Zn yield, i.e., the dotted line is for the 
metallicity-independent yield [GP(const)] and the dashed line is for 
the realistic metallicity-dependent yield [GP(var)]. 
(b) The corresponding non-LTE corrections used for deriving the 
[Zn/Fe] values shown in panel (a), plotted as functions of [Fe/H].
}
\end{figure}

\begin{figure}
  \begin{center}
    \FigureFile(155mm,155mm){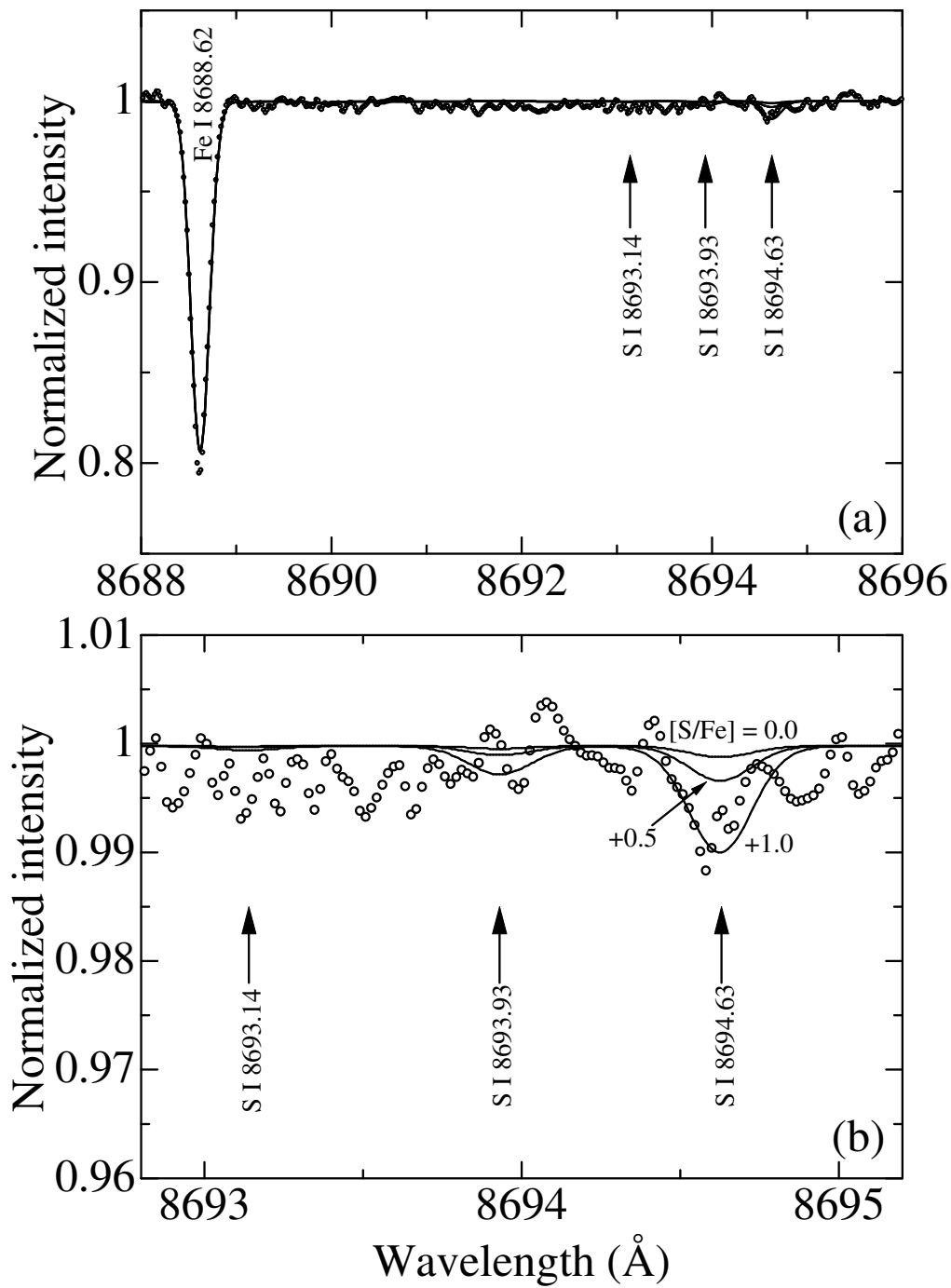}
  \end{center}
\caption{Open circles: Spectrum of HD~140283 observed with 
ESO/UVES, which was taken from the spectral database of Paranal 
Observatory  Project (Bagnulo et al. 2003). 
Solid lines: Theoretical spectra computed with the atmospheric 
parameters ($T_{\rm eff}$, $\log g$, $v_{\rm t}$, [Fe/H]) of 
(5690~K, 3.69, 1.5~km~s$^{-1}$, $-2.42$) for three sulfur 
abundances of [S/Fe] = 0.0, +0.5, and +1.0, where the calculation 
was done in LTE, because the non-LTE effect is negligibly small
for such very weak S~{\sc i} 8693--4 lines at this parameter range
(cf. electronic table E3). The computed spectra were 
convolved with a Gaussian function, adequately chosen 
to accomplish a good fit for the Fe~{\sc i} 8688.62 line.
(a) 8688--8696 $\rm\AA$ region for a wide view; (b) magnified 
8692.8--8695.2 $\rm\AA$ region for detailed inspection.}
\end{figure}

\end{document}